\documentclass[12pt]{article}

\usepackage[colorlinks=true,citecolor=blue, linkcolor=blue]{hyperref}

\usepackage{setspace}
\usepackage{authblk}
\usepackage[top=1in, bottom=1in, left=1in, right=1in]{geometry}

\usepackage{multirow}
\usepackage{enumitem}

\usepackage{amsmath}
\usepackage{graphicx}
\usepackage{natbib}
\usepackage{url} 
\usepackage{ulem}

\usepackage{tikz}
\usetikzlibrary{matrix,decorations.pathreplacing, calc, positioning,fit, arrows}
\usetikzlibrary{overlay-beamer-styles}
\usetikzlibrary{calligraphy}

\usepackage{tikz}
\usetikzlibrary{decorations.pathreplacing,calc}

\usepackage{mathtools}
\newcommand\upvertarrowbox[3][6ex]{%
  \begin{array}[b]{@{}c@{}} 
  \makebox[0pt]{\normalsize$\ensuremath{#3}$}\\ 
  \left\uparrow\vcenter{\hrule height #1}\right.\kern-\nulldelimiterspace\\
  #2
  \end{array}%
}

\newcommand\downvertarrowbox[3][6ex]{%
  \begin{array}[t]{@{}c@{}} #2 \\
  \left\downarrow\vcenter{\hrule height #1}\right.\kern-\nulldelimiterspace\\
  \makebox[0pt]{\normalsize$\ensuremath{#3}$} 
  \end{array}%
}

\newcommand\vertarrowboxx[4][6ex]{%
  \begin{array}[]{@{}c@{}} 
    \makebox[0pt]{\normalsize$\ensuremath{#3}$}\\ 
  \vcenter{\hrule height 0pt width 0pt depth #1}\\ 
  #2 \\ 
  \vcenter{\hrule height 0pt width 0pt depth #1}\\ 
  \makebox[0pt]{\normalsize$\ensuremath{#4}$} 
  \end{array}%
}

\newcounter{arrow}
\usepackage{booktabs} 

\usepackage{graphicx}
\usepackage{amsmath}
\usepackage{amssymb}
\usepackage{amsthm}
\usepackage{graphicx}
\usepackage{epstopdf}
\usepackage{comment}
\usepackage{subcaption}
\usepackage{esint}
\usepackage{bm}
\usepackage{latexsym}

\numberwithin{equation}{section}

\usepackage[linesnumbered,ruled,vlined]{algorithm2e}
\SetKwInput{kwInit}{Initialize}

\newcommand{\E}{\operatorname{E}}
\newcommand{\V}{\operatorname{var}}
\newcommand{\T}{\operatorname{T}}
\newcommand{\B}{\boldsymbol}

\newcommand{\OP}{ }
\newcommand{\bb}{\B{\OP{b}}}

\newcommand{\bx}{\bm{x}} 
\newcommand{\blambda}{\boldsymbol{\lambda}} 
\newcommand{\bphi}{\boldsymbol{\phi}} 
\newcommand{\bbeta}{\boldsymbol{\beta}}

\usepackage{xcolor}
\DeclareMathOperator*{\argmin}{arg\,min}

\newcommand{\bs}{\bm{s}}

\newcommand{\bh}{\bm{h}}
\def\trans{^{\rm T}}

\newcommand{\bI}{\mathbf{I}}
\newcommand{\bG}{\mathbf{G}}

\newcommand{\bX}{\boldsymbol{X}}

\newcommand{\bU}{\boldsymbol{U}}

\newcommand{\bS}{\mathbf{S}}

\newcommand{\bzeta}{\boldsymbol{\zeta}}

\newcommand{\bmu}{\boldsymbol{\mu}}

\newcommand{\bkappa}{\boldsymbol{\kappa}}
\newcommand{\bzero}{\mathbf{0}}

\def\E{E}

\def\wh{\hat}

\def\trans{^{\rm T}}

\newtheorem{theorem}{Theorem}[section]
\newtheorem{lemma}{Lemma}[section]

\newtheorem{corollary}{Corollary}[section]

\newtheorem{assumption}{Assumption}[section]

\title{Robust propensity score weighting estimation under missing at random}
\author[1]{Hengfang Wang}
\author[2]{Jae Kwang Kim \thanks{
E-mail addresses: hengfang@fjnu.edu.cn, jkim@iastate.edu, jshan411@gmail.com,  youngjomail@gmail.com}}
\author[3]{Jeongseop Han}
\author[4]{Youngjo Lee}
\affil[1]{School of Mathematics and Statistics \&  Fujian Provincial Key Laboratory of Statistics and Artificial Intelligence, Fujian Normal University, Fuzhou, Fujian, 350007, China}
\affil[2]{Department of Statistics, Iowa State University, Ames, IA, 50011, U.S.A.}
\affil[3]{Department of Statistics, Seoul National University, Seoul, 08826, Korea (South)}
\affil[4]{Department of Statistics, Seoul National University, Seoul, 08826, Korea (South)}
\date{}

\begin{document}

\maketitle

\begin{abstract}
	\label{section:abs}
	Missing data is frequently encountered in many areas of statistics. One popular approach to address this issue is through the use of propensity score weighting. However, correctly specifying the statistical model can be a daunting task.
	Doubly robust estimation is attractive, as the consistency of the estimator is guaranteed when either the outcome regression model or the propensity score model is correctly specified. 
	In this paper,  we first employ information projection to develop an efficient and doubly robust estimator via indirect model calibration. The resulting propensity score estimator can be equivalently expressed as a doubly robust regression imputation estimator by imposing the internal bias calibration condition in estimating the regression parameters. 
	In addition, using the $\gamma$-divergence measure, we generalize the information projection to allow for outlier-robust propensity score estimation.  
	The study includes the presentation of certain asymptotic properties and findings from a simulation study, which demonstrate that the proposed method enables robust inference, not only in cases of various model assumptions being violated but also in the presence of outliers. 
	A real-life application is also presented using data from the Conservation Effects Assessment Project.
\end{abstract}

\section{Introduction}
\label{section:RI_intro}
Missing data represents a 
 fundamental challenge in statistics.  Ignoring missing data can lead to biased estimates of parameters, loss of information, decreased statistical power, increased standard errors, and  
 weak generalizability of findings, as highlighted by  \cite{dong2013}. In addition, the missing data framework is very useful in defining problems in other research areas, such as causal inference or offline policy evaluation. 
Propensity score (PS) weighting is a popular method to handle missing data.  It often employs a response propensity model, but correct specification of the statistical model can be challenging in the presence of missing data. 
How to make the propensity score weighting method less dependent on the response propensity model is an important practical problem.

In the quest for robust PS  estimation, we find two distinct avenues. One approach embraces flexibility, employing nonparametric or semiparametric models to construct robust propensity scores.  The nonparametric kernel method \citep{hahn1998}, the  sieve logistic regression method \citep{hirano03}, the general calibration method of \cite{chan2016} using increasing dimension of the basis functions,   the generalized covariate balancing estimator using tailored loss functions \citep{zhao2019} and the random forest approach of \cite{dagdoug2021} are examples of the robust propensity score estimation method using flexible models. 
The other path, which we explore in this paper, leans on the outcome regression model explicitly to achieve doubly robust estimation.  
The literature is replete with investigations in this arena, with notable contributors like
 \cite{bang05}, \cite{cao09}, \cite{kimhaziza14}, \cite{han2013},  \cite{chenhaziza2017}, and 
\cite{yang2020}.

Our journey takes the latter route as we construct a unified framework for doubly robust PS  estimation under the setup of missing at random \citep{rubin1976}. To achieve the goal, we
apply the information projection \citep{csiszar2004} to the PS  weighting problem, while adhering to  covariate-balancing constraints. 
Specifically, the initial PS weights are constructed from the working PS model, but the balancing constraints, crucial for achieving double robustness in the PS weighting estimator, are derived from the working outcome regression model. This maneuver can be perceived as indirect model calibration. Information projection is used to obtain the augmented PS model.

 Remarkably, the  PS weighting estimator that is obtained from information projection  
 can also be cast as a special type of regression imputation estimator, incorporating balancing functions as covariates in the outcome regression model and the regression coefficients are computed using the weights computed from the final PS weights.  
 This algebraic equivalence, known as self-efficiency, forms the cornerstone of our journey toward outlier-robust  PS estimation. In practice, outliers do exist and the presence of outliers  can significantly undermine the efficiency of estimators. 
While   outlier-robust  procedures are well studied in the literature on regression or classification \citep{stefanski1986, wu2007, zhang2010}, their application in the context of missing data remains underexplored. 
To our best knowledge, the construction of an outlier-robust PS weighting  estimator has not been investigated in the literature. 


To fill in this important research gap, we embark on the creation of an outlier-robust regression imputation technique and subsequently align it with a PS weighting estimator by imposing the self-efficiency condition. In doing so, we set forth on a path previously untraveled in the literature - a journey toward constructing an outlier-robust PS  estimator.  
Specifically, the information projection approach to developing the augmented PS model and obtaining the doubly robust PS estimation is based on the Kullback-Leibler   divergence. 
The $\gamma$-power divergence, originally proposed by \cite{basu1998} and further developed by \cite{eguchi2021}, is a generalization of the
Kullback-Leibler divergence to expand the class of statistical models by introducing an additional
scale parameter $\gamma$. 
Furthermore, it is well known that the statistical model derived from the $\gamma$-power divergence produces robust inferences against outliers. We adopt
the $\gamma$-power divergence to develop an  outlier-robust  regression imputation  estimator. By self-efficiency, the regression estimator can be expressed as a PS weighting estimator. Therefore, we obtain an outlier-robust PS weighting  estimator.   The resulting estimator is also robust against misspecification of the response probability model. 



The structure of the paper is as follows. In Section \ref{section:RI_basic_setup},  the basic setup and the research problem are introduced. In Section~\ref{section: proposed_method}, we develop a balancing constraint to adjust propensity weights and construct an augmented propensity model using information projection. 
 In Section \ref{section:RI_info_proj}, we present the use of the $\gamma$-power divergence to enlarge the class of propensity score models and develop outlier-robust doubly robust estimators. 
 Results from two limited simulation studies are presented in Section \ref{section:RI_simulation}. 
 A real-life application using data from the
Conservation Effects Assessment Project is presented in Section~\ref{section: CEAP_application}. Some concluding remarks are made in Section~\ref{section: conclusion}. 
 All required proofs are presented in the Appendix.

 \section{Basic Setup}
 \label{section:RI_basic_setup} 
 
 Suppose that there are $n$ independently and identically distributed realizations of $(\bX, Y, \delta )$, denoted by $\{ (\bx_{i},y_{i},\delta_{i}):i=1,\ldots ,n \} $, where $\bx_{i}$ is a vector of observed covariates and $\delta_{i}$ is the missingness indicator associated with unit $i$. In particular, $\delta_{i} = 1$ if $y_{i}$ is observed and $\delta_{i} = 0$ otherwise. Thus, instead of observing $(\bx_i, y_i, \delta_i)$, we only observe $(\bx_i, \delta_i, \delta_i y_i)$ for $i=1, \ldots, n$. We are interested in estimating $\theta = E(Y)$ from the observed sample.

 Suppose that we have a working outcome regression (OR) model given by  
 \begin{equation}
 m_0\left ( \B{x}; \B{\beta}\right ) \in \mbox{span} \{\OP{b}_{0}(\B{x}),\OP{b}_1(\bx), \ldots, \OP{b}_L(\bx) \} 
 \label{outcome}
 \end{equation}
 for some given basis functions $\OP{b}_1(\bx), \ldots, \OP{b}_L(\bx)$ and  $\OP{b}_{0}(\B{x}) = 1$. The model (\ref{outcome}) can be expressed equivalently as 
 \begin{equation*}
	  y_i = m(\B{x}_{i}; \B{\beta}) + \varepsilon_i, ~ m(\B{x}_{i}; \B{\beta}) = \beta_0 + \beta_1 \OP{b}_1(\bx_i) + \ldots + \beta_L \OP{b}_L( \bx_i) 
 \end{equation*}
 for $\bbeta = (\beta_0, \beta_1, \ldots, \beta_L)^{\T}$, where $\varepsilon_i$ is an error term that is independent of $\bx_i$ and satisfies $\E(\varepsilon_i)=0$. 
 Furthermore, the assumption of missing at random \citep{rubin1976} implies that $\varepsilon_{i}$ and $\delta_{i}$ are conditionally independent given $\bx_{i}$.

 We are interested in using the following propensity score weighting (PSW)  estimator 
 \begin{equation} 
 \hat{\theta}_{\rm PSW} =\frac{1}{n} \sum_{i=1}^n \delta_i \omega_i y_i \label{ps}.
 \end{equation}  
 The following lemma provides a sufficient condition for the unbiasedness of the propensity score weighting estimators under the model (\ref{outcome}). 
 \begin{lemma}
	 Assume that the response mechanism is missing at random. If the  weights $\omega_i$'s satisfy 
 \begin{equation} 
 \sum_{i=1}^n \delta_i \omega_i \left[ \OP{b}_{0} (\B{x}_{i}),\OP{b}_{1} (\B{x}_{i}), \ldots, 
 \OP{b}_{L} (\B{x}_{i}) \right] = \sum_{i=1}^n \left[ \OP{b}_{0} (\B{x}_{i}),\OP{b}_{1} (\B{x}_{i}), \ldots, 
 \OP{b}_{L} (\B{x}_{i}) \right] , 
 \label{balancing}
 \end{equation}
  then $\hat{\theta}_{\rm PSW}$ in \eqref{ps} is unbiased for $\theta$ under the OR  model in (\ref{outcome}). 
 \label{lemma1}
 \end{lemma}
 
 By Lemma \ref{lemma1}, condition (\ref{balancing}) is the key condition that incorporates the OR  model into the PSW estimator. Condition (\ref{balancing}) is often called the covariate balancing condition \citep{imai2014} in the missing data literature. It is closely related to the calibration estimation in survey sampling \citep{deville1992calibration}.  Under the OR model in (\ref{outcome}), the covariate-balancing condition in (\ref{balancing}) implies that  
 \begin{equation}
  \sum_{i=1}^n \delta_i \omega_i\E(Y_i \mid \bx_i) =\sum_{i=1}^n \E(Y_i \mid \bx_i), \label{mcal}
  \end{equation}
  which is often referred to as the   model calibration \citep{wu2001}.  
 To distinguish from the direct model calibration of \cite{wu2001}, we may call (\ref{balancing}) the \emph{indirect model calibration}. 
 
 The indirect model calibration condition also implies the following algebraic equivalence. 
 \begin{lemma}
	If the weights $\omega_i$ satisfy (\ref{balancing}), then we have 
 \begin{equation} 
 \sum_{i=1}^n \delta_i \omega_i y_i = \sum_{i=1}^n \{ \delta_i y_i + (1- \delta_i) \B{\OP{b}}_{i}^{\T} \hat{\B{\beta}}   \} , 
 \label{eq7} 
 \end{equation} 
	where $\hat{\bbeta}$ satisfies 
 \begin{equation}
 \sum_{i=1}^n \delta_i \left( \omega_i -1  \right)  \left ( y_i - \B{\OP{b}}_{i}^{\T} \hat{\B{\beta}} \right )=0 .
 \label{ibc-2}
 \end{equation}
 and $\bb_{i}= ( 1, \OP{b}_1(\bx_{i}), \ldots, \OP{b}_L(\bx_{i}) )^{\T}$. 
 \label{lem2}
 \end{lemma}

 Lemma \ref{lem2} means that, under (\ref{ibc-2}),  the PSW  estimator that satisfies the covariate-balancing condition is algebraically equivalent to a regression imputation estimator using the balancing functions as covariates in the outcome regression model. In other words, regression imputation under the outcome regression model in (\ref{outcome}) can be viewed as a PSW  estimator where the propensity score weights  $\omega_i$ and the estimated regression coefficients $\hat{\bbeta}$ are related by equation (\ref{ibc-2}). 
 Equation (\ref{eq7}) means that the final propensity score weights ${\omega}_i$ do not directly use the outcome regression model (\ref{outcome}) for imputation, but implement regression imputation indirectly. Condition (\ref{eq7}) is referred to as the self-efficiency condition.

 We now wish to achieve double robustness by including the PS  model. Suppose that we have the following working PS  model 
 $$ 
P \left( \delta = 1 \mid \bx \right) = \pi_1 ( \bx; \bphi_0)
 $$
 for some $\bphi_0$ with a known function $\pi_1(\cdot) \in (0,1)$. 
An example is the logistic regression model such as $\mbox{logit} \{ \pi_1(\bx; \bphi_0) \} = \bx \trans  \bphi_0$.  
 Under the working propensity score model, the PS  weights 
 are computed as
 $\omega_i =\{ \pi_1(\bx_i ; \hat{\bphi} ) \}^{-1}:= \hat{d}_i$, where $\hat{\bphi}$ is the maximum likelihood estimator (MLE)  of $\bphi$. However, the PS  weight $\hat{d}_i$ does not necessarily satisfy the balancing condition in (\ref{balancing}). Therefore, to reduce the bias due to model misspecification in the propensity score model, it makes sense to impose the balancing condition in the final weighting. To achieve this goal, \cite{hainmueller2012} proposed the so-called entropy balancing method that minimizes 
 \begin{equation}
  Q(\B{\omega}) = \sum_{i=1}^n \delta_i \omega_i\log ( \omega_i/ \hat{d}_i ), \label{entropy} 
 \end{equation}
 subject to the balancing constraint in (\ref{balancing}). 
 \cite{chan2016} generalized this idea further to develop a general calibration weighting method that satisfies the covariance balancing property with increasing dimensions of basis functions $\B{\OP{b}} ( \bx)$.  However, the choice of distance measure is somewhat unclear. In the next section, we adopt  the information projection approach to develop a unified approach to  doubly robust propensity score weighting.

 \section{Proposed method}
 \label{section: proposed_method}

 We now develop an alternative approach to 
 modifying the initial propensity score weights to satisfy the covariate balancing condition in (\ref{balancing}). The proposed approach consists of two parts. One is the modeling part  and the other is the parameter estimation part. In the modeling part, we use information projection innovatively to obtain the propensity score model incorporating the moment constraints from the outcome regression model. 
  Information projection is a powerful tool for incorporating the moment constraints into the final model.  
 
 \subsection{Information projection}

 Instead of using a distance measure for propensity weights, as in \eqref{entropy}, we use the Kullback-Leibler  divergence measure properly  to develop the information projection under some moment constraint obtained from the outcome regression model. 
 Because the Kullback-Leibler divergence is defined in terms of density functions, we need to formulate the weighting problem as an optimization problem for density functions.

 To apply the information projection, using the Bayes theorem, we obtain 
 $$ \frac{ \OP{P} ( \delta =0 \mid \B{x}) }{\OP{P} ( \delta =1 \mid \B{x})  } = \frac{1-p}{ p } \times  \frac{ f_0( \B{x})}{ f_1( \B{x})} ,$$
 where $p = \OP{P}(\delta = 1)$, and $f_k( \B{x}) = f( \bx \mid \delta=k)$ is the density function for $\bx$ given $\delta = k$ for $k=0,1$.   Thus, the propensity score (PS) weight can be written as 
 \begin{equation}
 \omega(\bx) \equiv \frac{1}{\OP{P} ( \delta =1 \mid \B{x}) } 
 = 1+ c \times \frac{ f_0( \B{x})}{ f_1( \B{x})}. 
 \label{dr1}
 \end{equation}
 where $c=1/p - 1$. For a fixed $f_1$, the propensity weight function $\omega(\bx)$ is completely determined by $f_0$.
 Furthermore, assume that the basis function ${\bb}(\bx)$ in the outcome regression model in (\ref{outcome}) is integrable in the sense that
 \begin{equation} 
  p \E_1\{ \B{\OP{b}} (\B{X}) \} + (1-p) \E_0 \{ \B{\OP{b}} ( \B{X}) \} = \E \{ \B{\OP{b}}(\B{X}) \} , 
 \label{moment}
 \end{equation} 
 where $\E_k \{ \B{\OP{b}} ( \B{X}) \} = \int \B{\OP{b}} (\bx) f_k( \bx) d \mu(\bx)$ for $k=0,1$ and $\mu$ is the Lebesgue measure. Thus, for a given $f_1$, equation  (\ref{moment}) can serve as a constraint on $f_0$ when   $\E\{ \B{\OP{b}} (\B{X}) \} $ is known. 
 
 Let $\pi_1^{(0)}(\bx)$ be the PS function corresponding to  a  ``working'' model for $\pi_1(\bx) = \OP{P}(\delta = 1\mid \bx)$.  
 For a fixed $f_1$, by (\ref{dr1}), we can define $f_0^{(0)}$ to satisfy 
 \begin{equation}
 \frac{1}{
 \pi_1^{(0)}(\bx)} = 1+ c \times \frac{ f_0^{(0)} (\bx) }{ f_1(\bx) }. 
 \label{dr5}
 \end{equation}
 We can treat $f_0^{(0)}$ as the baseline density for $f_0$ derived from the working PS function $\pi_1^{(0)}(\bx)$. Our objective is to modify $f_0^{(0)}$ to satisfy (\ref{moment}).  
 Finding the density function $f_0$ satisfying the balancing condition in (\ref{moment}) can be formulated as an optimization problem using the Kullback-Leibler  divergence. Given the baseline density $f_0^{(0)}$, the Kullback-Leibler divergence of $f_0$ at $f_0^{(0)}$ is defined by 
 \begin{equation} 
 D_{\rm KL} \left ( f_0 \|  f_0^{(0)} \right ) 
 = \int \log \left( \frac{ f_0}{f_0^{(0)}} \right) f_0  d \mu .  \label{kl}
 \end{equation} 
 Given $f_0^{(0)}$, we wish to find the minimizer of 
 $ D_{\rm KL} ( f_0 \|  f_0^{(0)}) $
 subject to (\ref{moment}). 
 The Lagrangian function can be formulated as
 \begin{align}\label{Lagrangian}
    \mathcal{L}(f_{0}, \blambda)  = &
   \int \log \left( f_0/f_0^{(0)} \right) f_0  d \mu\notag\\
  & +
    \blambda\trans \left[p
    \int \bb(\bx) f_{1}( \bx )d\mu
     + (1-p) 
      \int \bb(\bx) f_{0}( \bx )d\mu
     - \E \{ \B{\OP{b}}(\B{X}) \}\right],
 \end{align}
 where $\B{\lambda}$ is the Lagrange multiplier.
As $f_{1}$ is fixed and $\E \{ \B{\OP{b}}(\B{X}) \}$ is known, by taking the derivative with respect to $f_{0}$ in \eqref{Lagrangian}, one may arrive at $ f_0^* ( \B{x}; \blambda) \propto f_0^{(0)} ( \B{x}) \exp \{ \B{\OP{b}}\trans(\B{x})\B{\lambda} \}$. Furthermore, the density constraint $\int f_0^* ( \B{x}; \blambda) d\bx = 1$ leads to  
 \begin{equation}
 f_0^* ( \B{x}; \blambda) = f_0^{(0)} ( \B{x})  \frac{ \exp \{ \B{\OP{b}}\trans(\B{x})\B{\lambda} \} }{ \int \exp \{ \B{\OP{b}}\trans(\B{x}) \B{\lambda} \}  f_0^{(0)} ( \B{x})  d \mu (\bx) },
 \label{8a}
 \end{equation}
 with abuse of notation for $\blambda$.  By (\ref{8a}),  we obtain 
 \begin{equation}
  \frac{ f_0^* ( \B{x}; \blambda)}{f_1 (\B{x})  } = \frac{ f_0^{(0)} ( \B{x})}{f_1 (\B{x})  } \times \frac{ \exp \{  \sum_{j=1}^L {\lambda}_j \OP{b}_j (\B{x}) \}}{ \int \exp \{  \sum_{j=1}^L {\lambda}_j \OP{b}_j (\B{x}) \} f_0^{(0)} (\bx) d \mu(\bx)}. 
  \label{10} 
 \end{equation}
 Combining the above results, we can establish the following lemma. 
 \begin{lemma} 
 Given the  PS function  $\pi_1^{(0)}(\bx)$ from the working PS  model, the  final PS   function  minimizing the Kullback-Leibler divergence in (\ref{kl}) subject to the balancing condition (\ref{moment}) is given by 
	 \begin{equation}\pi_1^* ( \bx;  \blambda) = \frac{{\pi}_1^{(0)}(\bx) }{{\pi}_1^{(0)} (\bx) +   \{1- {\pi}_1^{(0)} (\bx) \} \exp \left\{\lambda_0 + \sum_{j=1}^L {\lambda}_j {b}_{j} (\bx) \right\}  },\label{augps} \end{equation}
 where $\lambda_0, \lambda_1, \ldots, \lambda_L$ are the Lagrange multipliers satisfying (\ref{moment}). 
 \end{lemma}

 In (\ref{augps}), the Lagrange multiplier $\blambda$ is determined to satisfy  the balancing condition (\ref{moment}) with 
 \begin{align*}
 \E_0 \{ \B{\OP{b}} ( \B{X}) \} 
 =   \frac{ \int \B{\OP{b}} (\bx)O^{(0)}(\bx)\exp \{  \sum_{j=1}^L {\lambda}_j \OP{b}_j (\B{x}) \} f_1 (\bx)  d \mu( \bx)
 }{ \int  O^{(0)}(\bx)\exp \{  \sum_{j=1}^L {\lambda}_j \OP{b}_j (\B{x}) \} f_1 (\bx) d \mu(\bx)}. \end{align*} 
 where $O^{(0)}(\bx) = \{ \pi_1^{(0)}(\bx) \}^{-1} -1$. 
 If the working propensity score model is correct, then the balancing constraint (\ref{moment}) is already satisfied. In this case, we have $\lambda_j \equiv 0$ for all $j= 1, \ldots, L$. Thus, the information projection is used to obtain the augmented propensity score model \citep{kimriddles12}.

 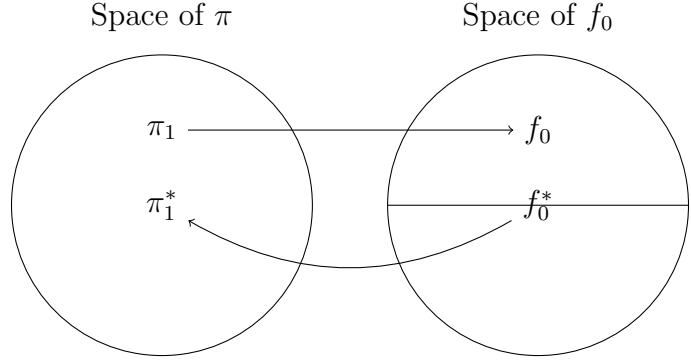
\begin{figure} 
  \begin{center} 
  \begin{tikzpicture}
   
  \draw (-2.5,0) circle (2cm); 
  \draw (2.5,0) circle (2cm); 
  \draw (0.5,0) -- (4.5, 0) ;
      \node (H) at (-2.5,2.5) {Space of $\pi$};

  \node (a) at (-2.5, 1.0) {$\pi_1$}; 
  
  \node (b) at (-2.5, 0.0) {$\pi_1^*$};

  \node (c) at (2.5, 1.0) {$f_0$}; 
  
  \node (d) at (2.5, 0.0) {$f_0^*$}; 
      \node (HF) at (2.5,2.5) {Space of $f_0$};

  \draw[->] (a) -- (c);
  
  \draw[->] (d) edge[bend left] (b);
  
  \end{tikzpicture}
  \caption{A graphical illustration of obtaining $\pi_1^*(x)$ in (\protect\ref{augps}) using the information projection for $f_0$, where the line running halfway across the right circle represents the space of $f_1$ satisfying the covariate-balancing constraints.}
   \label{fig-1}

   \end{center} 
  
  \end{figure}

   Figure \ref{fig-1} presents a graphical summary of obtaining the augmented PS model in (\ref{augps}). Using the relationship in (\ref{dr1}), we can transform $\pi_{1}(x)$ to $f_0(x)$ and then apply the information projection to obtain $f_0^*(x)$ in (\ref{8a}), which in turn finds $\pi_{1}^*(x)$ in (\ref{augps}) by applying (\ref{dr1}) again. 
  
In practice,  the base functions for the ``working'' outcome model are chosen by standard model selection techniques. For example, \cite{shortreed2017outcome} used adaptive Lasso to select the covariates in the outcome regression model.  \subsection{Model parameter estimation}
   \label{section:model_parameter_estimation}
   
   We now discuss parameter estimation in the augmented propensity score model in (\ref{augps}). Suppose that the working propensity score model is given by $\pi_{1}^{(0)}( \bx; \bphi)$ with the unknown parameter $\bphi$. We can estimate $\bphi$ by maximizing 
   $$ \ell ( \bphi) = \sum_{i=1}^n \left[ \delta_i \log \left \{ \pi_{1}^{(0)}( \B{x}_i; \bphi) \right \} + ( 1- \delta_i) \log \{ 1- \pi_{1}^{(0)} (\B{x}_i; \bphi) \} \right] $$
   with respect to $\bphi$. Note that the propensity score model has nothing to do with the OR  model in (\ref{outcome}). 
   
   Now, to incorporate the OR  model in (\ref{outcome}), 
   we use the augmented propensity score model in (\ref{augps}) to obtain the final propensity score weights 
   \begin{equation}
   \hat{\omega}_{i} = 1+  ( \hat{d}_i -1 ) \exp \left\{  \sum_{j=0}^L \hat{\lambda}_j \OP{b}_{j} (\B{x}_i) \right\} , 
   \label{wgt2}
   \end{equation} 
    where $\hat{d}_i = \{ \pi_{1}^{(0)}( \B{x}_i; \hat{\bphi})\}^{-1}$ and  $\hat{\lambda}_0, \hat{\lambda}_1, \ldots, \hat{\lambda}_L$ are computed from the calibration equation:  
    \begin{equation} 
   \sum_{i=1}^n \delta_i \hat{\omega}_{i}\OP{b}_{j}(\B{x}_{i}) = \sum_{i=1}^n \OP{b}_{j}(\B{x}_{i}), \ \ \ \forall j=0,1, \ldots, L.
   \label{calib3}
   \end{equation} 
  By construction, the PS weights in (\ref{wgt2}) satisfies the covariate-balancing property  on the basis functions in the outcome regression model in (\ref{outcome}). By Lemma 2.1, the covariate balancing property implies unbiasedness of the PSW estimator under the OR model. On the other hand, if the working PS model is correct, then $\hat{\lambda}_j \rightarrow 0$ in probability as $n\rightarrow \infty$ for $j=0,1, \ldots, L$. In this case, $\hat{\omega}_i$ will converge to $\hat{d}_i$ and 
   the PSW estimator is consistent under the PS model.

   As discussed in Section~\ref{section:RI_basic_setup}, the propensity score estimator satisfying the indirect model calibration condition can be expressed as a regression imputation estimator. Since $\hat{\omega}_i$ satisfy (\ref{calib3}), by Lemma \ref{lem2}, we can obtain 
   \begin{equation}
   \frac{1}{n} \sum_{i =1}^n \delta_i \hat{\omega}_i y_i = \frac{1}{n} \sum_{i=1}^n \left\{ \delta_i y_i + (1- \delta_i) \hat{y}_i \right\}  ,
   \label{self}
   \end{equation}
   where $\hat{y}_i = \B{\OP{b}}_i^{\T} \hat{\bbeta}$ and
   \begin{equation}
   \hat{\bbeta} = \left\{
   \sum_{i=1}^n \delta_i \left(\hat{\omega}_i -1 \right) \B{\OP{b}}_i \B{\OP{b}}_i^{\T}
   \right\}^{-1} \left \{ \sum_{i=1}^n \delta_i \left(\hat{\omega}_i -1 \right) \B{\OP{b}}_i y_i \right \} .\label{ibc2}
   \end{equation}

   Note that, by (\ref{wgt2}), $\hat{\bbeta}$ in (\ref{ibc2})  can be expressed as the minimizer of 
    \begin{equation}
    Q (\bbeta) =  \sum_{i=1}^n \delta_i   \left( y_i - \bb_i^{\T} \bbeta\right)^2 ( \hat{d}_i -1) \hat{g}_i, \label{eq:Q_KL}
   \end{equation}
   where $\hat{g}_i = \exp ( {\bb}_i^{\T} \hat{\blambda})$.
   The first term $\hat{d}_i -1$ is the adjustment term to incorporate the working PS model into the regression imputation. 
   The second term $\hat{g}_i=\exp (   {{\bb}}_i^{\T} \hat{{\blambda}}  )$ is to achieve covariate balance in the PS weights, which is a sufficient condition for self-efficiency.

  Now, if $g_i=1$ is used in (\ref{eq:Q_KL}), then the self-efficiency in (\ref{self}) will  not be satisfied. Instead, we only obtain 
  \begin{equation}
   \frac{1}{n} \sum_{i=1}^n \left\{ \delta_i y_i + (1- \delta_i) \hat{y}_i \right\} = \frac{1}{n} \sum_{i=1}^n \left\{ \hat{y}_i + \delta_i \hat{d}_i \left( y_i- \hat{y}_i\right) \right\} .
   \label{ibc}
   \end{equation}
  Condition (\ref{ibc}) is called the \emph{internal bias calibration} (IBC), which was originally termed by  \cite{firth1998} in the context of design-consistent estimation of model parameters under complex sampling. The imputation estimator  satisfying the IBC  condition (\ref{ibc}) achieves consistency even when the outcome regression model is incorrect, as long as the working PS model is correct. Thus, the  IBC  condition is a sufficient condition for the double robustness of the regression imputation estimator.

  For the choice of $\hat{g}_i = \exp ( {\bb}_i^{\T} \hat{\blambda})$, under the PS model, we obtain $\hat{g}_i \rightarrow 1$ in probability as $n\rightarrow \infty$. Thus, the IBC condition in (\ref{ibc}), or double robustness of the regression imputation estimator,  holds approximately.

   Equation (\ref{self}) deserves further discussion.  In the PSW  estimation, for a given $\hat{\bphi}$, the final estimator $\hat{\theta}_{\mathrm{PSW}}$ is a function of estimated nuisance parameter $\hat{\blambda}$ while the regression imputation estimator is a function of other estimated nuisance parameter   $\hat{\bbeta}$. Thus, $\hat{\blambda}$ and $\hat{\bbeta}$ are in one-to-one correspondence with each other  through the following estimating equation: 
   \begin{equation}
    \sum_{i=1}^n \delta_i ( y_i -\B{\OP{b}}_i^{\T} \hat{\bbeta} ) \B{\OP{b}}_i (\hat{d}_i -1 ) \exp (  \B{\OP{b}}_i^{\T}  \hat{\B{\lambda}} )
   = \B{0}. 
   \label{cond8}
   \end{equation}

  In summary, the proposed method can be implemented in the following steps. 
  \begin{enumerate}
  \item Specify a working PS model $\pi_{1}^{(0)}(\bx; \bphi)$ and estimate $\bphi$ using the ML estimation method to get $\hat{d}_i = \{ \pi_{1}^{(0)} (\bx; \hat{\bphi}) \}^{-1}$. 
  \item Specify a working OR model in (\ref{outcome}) to construct the augmented PS model in (\ref{augps}). 
  \item Compute $\hat{\blambda}$  from the calibration equation in (\ref{calib3}) to obtain $\hat{\omega}_i$ in (\ref{wgt2}). 
  \item Also, compute $\hat{\bbeta}$ from (\ref{ibc2}) to obtain the regression imputation estimator in (\ref{self}).   
  \end{enumerate}

  \section{Statistical properties}

 Now we formally describe the asymptotic properties of the augmented PSW  estimator using the final propensity score weight (\ref{wgt2}) with the calibration constraint in (\ref{calib3}). By the algebraic equivalence established in Lemma \ref{lem2}, the result is directly applicable to the regression imputation estimator using the regression coefficient in (\ref{ibc2}).

 Let $\pi_{1}^{(0)}(\B{x}; \B{\phi})$ be the working propensity score model, where $\B{\phi} \in \mathbb{R}^{p}$. We can estimate $\bphi$ by solving 
 \begin{equation}
 \bU_{1, n} ( \bphi) \equiv \frac{1}{n} \sum_{i=1}^n \left\{ \frac{\delta_i}{\pi_{1}^{(0)}( \bx_i; \bphi) } - 1 \right\} \bh(\bx_i; \bphi) = \B{0} 
 \label{ee5}
 \end{equation}
 for some $\bh(\bx; \bphi)$ such that the solution to (\ref{ee5}) exists uniquely almost everywhere. The estimating equation (\ref{ee5}) for $\bphi$ includes the score equation for $\bphi$ as a special case.
 For a given $\hat{\bphi}$, let $\hat{\bU}_2(\blambda \mid \hat{\bphi})$ be the estimating equation for ${\blambda}$. By (\ref{balancing}), we can estimate $\blambda$ by solving 
 \begin{equation} 
  \bU_{2, n}(\blambda \mid \hat{\bphi} ) \equiv \frac{1}{n} \sum_{i=1}^n \delta_i  \omega ( \B{x}_{i}; \hat{\bphi}, \blambda )  {\bb}_i  - \frac{1}{n}  
 \sum_{i=1}^n {\bb}_i= \mathbf{0}  , 
 \label{est3}
 \end{equation} 
 where 
 \begin{equation}
 \omega \left ( \B{x}_{i}; {\bphi}, {\blambda} \right ) = 1+ \left\{ \frac{1}{ \pi_{1}^{(0)} ( \bx_i ; {\bphi} )} -1    \right\} \exp \left ( \bb_{i}^{\T} \blambda \right ).
 \label{cond9-1}
 \end{equation}

 Further, define 
 \begin{align*}
\bG_{n}(\bphi) 
=& \frac{1}{n}\sum_{i=1}^{n}
\left[
\left\{1 - \frac{\delta_{i}}{\pi_{1}^{(0)}(\bx_{i};\bphi)}\right\} \left\{\frac{\partial \bh(\bx_{i}; \bphi)}{\partial \bphi}\right\}\trans \notag \right.\\
&\left. \qquad\qquad+  \frac{\delta_{i}}{ \{\pi_{1}^{(0)} (\bx_{i};\bphi)\}^{2}}
\frac{\partial \pi_{1}^{(0)}(\bx_{i};\bphi)}{\partial \bphi}\bh\trans(\bx_{i}; \bphi)
\right].
 \end{align*}
 We have the following assumptions before the statement of our main theorem.

\begin{assumption}\label{unique solution0}
The estimating equation  $\bU_{1, n}(\bphi) = \bzero$ has a unique solution $\wh{\bphi}$ and there exists a function $\bU_{1}(\bphi)$ such that $\bU_{1, n}(\bphi) \rightarrow \bU_{1}(\bphi)$ uniformly as $n \rightarrow \infty$ and $\bU_{1}(\bphi) = \bzero$ has a unique solution $\bphi^{\ast}$.  Further, The estimating equation  $\bU_{2, n}(\blambda \mid \bphi^*) = \bzero$ has a unique solution $\wh{\blambda}$ and there exists a function $\bU_{2}(\blambda\mid \bphi^*)$ such that $\bU_{2, n}(\blambda \mid \bphi^*) \rightarrow \bU_{2}(\blambda \mid \bphi^*)$ uniformly as $n \rightarrow \infty$ and $\bU_{2}(\blambda \mid \bphi^*) = \bzero$ has a unique solution $\blambda^{\ast}$.  
\end{assumption}

	\begin{assumption}\label{unique solution1}
The limiting function $\bU_{1}(\bphi)$ is differentiable and $\partial_{\bphi}\bU_{1}(\bphi)$ is continuous on a compact set $\mathcal{G}_{1}$ containing $\bphi^{\ast}$. In addition, the limiting function $\bU_{2}(\bphi\mid \bphi^*)$ is differentiable and $\partial_{\blambda}\bU_{2}(\blambda\mid\bphi^*)$ is continuous on a compact set $\mathcal{G}_{2}$ containing $\blambda^{\ast}$.
\end{assumption}

\begin{assumption}\label{unique solution2}
 The matrices $ \partial_{\bphi} \bU_{1}(\bphi)  \mid_{\bphi = \bphi^{\star}}$
	and $ \partial_{\blambda} \bU_{2}(\blambda\mid \bphi^{\ast}) 
		\mid_{\blambda = \blambda^{\ast}}$ are both 
		non-singular.
  \end{assumption}

\begin{assumption}\label{existence}
 There exists a function $\bG(\bphi)$ such that $\bG_{n}(\bphi) \rightarrow \bG(\bphi)$ uniformly as $n \rightarrow \infty$ and $\bG(\bphi^*)$ is non-singular. 
 \end{assumption}

Assumptions~\ref{unique solution0} -- \ref{unique solution2} are commonly used in estimating  equation theory and also ensure the existence of $\blambda^*$ and $\bphi^*$; see \cite{newey1994large} and \cite{kim2009unified} for more details.
 Note that $\blambda^*$ and $\bphi^*$ are  the probability limits of $\hat{\blambda}$ and $\hat{\bphi}$, respectively.  
Assumption~\ref{unique solution2} also guarantees the existence of $\bbeta^{*}$, which is the probability limit of $\hat{\bbeta}$ in  \eqref{ibc2}. Assumption~\ref{existence} ensures the existence of the nuisance parameter $\bkappa^*$ defined in the following Theorem~\ref{T1}, which helps us represent the influence function for the proposed estimator. 
If the maximum likelihood estimation is used to estimate $\bphi$, then $\bphi^*$ can be understood as the minimizer of the cross-entropy  
$$H( \pi_{1}, \pi_1^{(0)}(\bphi) ) =-  \E\left[ \pi_{1}(\bx) \log \left \{ \pi_{1}^{(0)} ( \bx; \bphi) \right \} +\{1- \pi_{1}(\bx) \} \log \left\{ 1- \pi_{1}^{(0)} (\bx; \bphi) \right\} \right]$$
 with respect to $\bphi$, 
 where $\pi_{1}( \bx)= \OP{P} ( \delta =1 \mid \bx)$ is the true response probability. 
 Now, using (\ref{cond9-1}),  we can treat 
 \begin{equation}
 \hat{\theta}_{\rm APSW} = \frac{1}{n} \sum_{i=1}^n \delta_i \omega ( \B{x}_{i};  \hat{\bphi},  \hat{\blambda} ) y_i: = \hat{\theta}_{\rm APSW}  ( \hat{\bphi}, \hat{\blambda} ) 
 \label{aps}
 \end{equation}
 as a function of $(\hat{\bphi}, \hat{\blambda})$ and 
 apply the standard Taylor linearization to obtain the following theorem, whose proof is presented in the Appendix.

 \begin{theorem}\label{T1}
  Let $\hat{\theta}_{\mathrm{APSW}}$ in (\ref{aps}) be the PSW   estimator under the augmented PS  model in (\ref{augps}) with $\hat{\bphi}$ and $\hat{\blambda}$ satisfying (\ref{ee5}) and (\ref{est3}), respectively. 
  Under the regularity conditions described in the Appendix, we have
  \begin{align}
   \hat{\theta}_{\rm APSW}  - \theta 
   = 
   \frac{1}{n} \sum_{i=1}^n  {\eta} (\bx_i, y_i, \delta_i) + o_p(n^{-1/2}), \label{33} 
  \end{align}
  where 
  \begin{align}
  {\eta} (\bx_i, y_i, \delta_i) = &\bb_i^{\T} {\bbeta}^* - \theta 
   + \delta_i \omega (\bx_i; \B{\phi}^*, \B{\lambda}^* ) \left( y_i - \bb_i^{\T} {\bbeta}^* \right) \notag\\
   &\qquad +  
  \left \{ 1- \frac{\delta_i}{\pi_{1}^{(0)} ( \bx_i; \bphi^*)}  \right \} \bh_i^{\T} \B{\kappa}^*  , 
  \label{influence}
  \end{align}
  $\bh_{i} = \bh(\bx_i; \bphi)$, $\omega(\bx; \B{\phi}, \B{\lambda})$ is defined in (\ref{cond9-1}), 
   and $\B{\kappa}^*$ is the probability limit of $\hat{\B{\kappa}}$ that satisfies 
  \begin{equation}
  \sum_{i=1}^n \delta_i (\hat{\pi}_{1, i}^{-1}-1) \left\{ \exp(\bb_i^{\T} \hat{\blambda}) ( y_i - \bb_i^{\T} \hat{\bbeta}) - \bh_i^{\T} \hat{\B{\kappa}} 
  \right\} \left [ \frac{\partial}{\partial \bphi}\mathrm{logit} \left \{\pi_{1}^{(0)}  \left (\bx_i; \bphi \right ) \right \} \Big |_{\bphi = \hat{\bphi}} \right ] = \B{0},
  \label{28}
  \end{equation}
 {where $\hat{\pi}_{1, i} = \pi_{1}^{(0)}(\bx_{i}; \wh{\bphi})$.}
  \label{theorem1}
  \end{theorem}
  
  In (\ref{influence}), $\eta( \bx, y, \delta)$ is called the influence function of $\hat{\theta}_{\rm APSW}$ \citep{tsiatis2006}. 
  Note that we can obtain 
  \begin{align*}
  \E\left\{ \eta(\B{X}, Y, \delta) \right\}=&  \E\left\{ \delta \omega(\B{X}; \bphi^*, \blambda^*) Y \right\}- \theta \\
  &+ \E\left[ \{ 1 - \delta  \omega (\B{X}; \B{\phi}^*, \B{\lambda}^*) \} \bb^{\T}(\B{X}) \bbeta^* \right] \\
  &+ \E \left[ \left\{ 1- \frac{\delta}{\pi_{1}^{(0)} ( \B{X}; \bphi^*)} \right\} \bh^{\T} (\B{X}) \B{\kappa}^* 
  \right],
  \end{align*}
  where the second term is equal to zero by the definition of $\blambda^*$ and the third term is equal to zero by the definition of $\bphi^*$. 
  If the outcome regression model is correctly specified, we have 
  \begin{align*} 
  \E\left\{ \delta \omega(\B{X}; \bphi^*, \blambda^*) Y \right\} =& \E\left\{ \delta \omega(\B{X}; \bphi^*, \blambda^*) \B{b}\trans(\B{X}) \bbeta^* \right\} \\=&
  \E\left\{  \B{b}\trans(\B{X}) \bbeta^* \right\} = \E(Y),
  \end{align*} 
  where the second equality follows from the definition of $\blambda^*$.  Furthermore, the correct specification of the outcome regression model gives ${\bkappa}^*= \bzero$. Thus, we can summarize the asymptotic result in the outcome regression model as follows. 
  \begin{corollary}
    Suppose that the regularity conditions in Theorem~\ref{T1} hold and the outcome regression model is correctly specified. Then,  
   \begin{equation}
     \sqrt{n}\left( \hat{\theta}_{\rm APSW} - \theta  \right) \stackrel{\mathcal{L}}{\longrightarrow} N(0,V_{\rm OR} ),
     \label{result0}
   \end{equation}
   where 
   \begin{equation} 
   V_{\rm OR} =  \V \left\{ \E(Y \mid \B{X} ) \right\} + \E\left[ \delta \{ \omega( \B{X}; \bphi^*, \blambda^*) \}^2 \V \left (Y \mid \B{X} \right ) \right]. 
  \label{var_KL}
  \end{equation} \end{corollary}

  Now, consider the case where the propensity score model is correctly specified. In this case, we obtain $\blambda^*=\bzero$ and  $\pi_{1}^{(0)} ( \bx; \bphi^*) = \OP{P}( \delta =1 \mid \bx )$. Thus, 
  \begin{equation*} 
  \E\left\{ \delta \omega(\B{X}; \bphi^*, \blambda^*) Y \right\} = \E\left[\OP{P}(\delta = 1 \mid \B{X}) \{ \pi_{1}^{(0)} ( \B{X}; \bphi^*) \}^{-1}  Y \right] = \E(Y).
  \end{equation*} 
  Therefore, we can obtain the following result. 
  \begin{corollary}
  Suppose that the regularity conditions in Theorem~\ref{T1} hold and the  propensity score  model is correctly specified. Then,  
   \begin{equation}
     \sqrt{n}\left( \hat{\theta}_{\rm APSW} - \theta  \right) \stackrel{\mathcal{L}}{\longrightarrow} N(0,V_{\rm PS}   ),
     \label{result2}
   \end{equation}
  where  
  \begin{align*} 
  V_{\rm PS} 
  =& \V \left( Y  \right)  +
  \E\left[ \left\{ \frac{1}{\pi_{1}(\bX) }-1 \right\} \{ Y -  \bb^{\T}(\B{X}) \bbeta^* - \bh^{\T} (\B{X}) \B{\kappa}^* \}^2  \right]
  \end{align*}
  and $\pi_{1}(\bX) = P( \delta =1 \mid \bX)$. 
  \end{corollary}
  Thus, the efficiency gain using the estimated propensity score function over the true one is visible only when the PS model  is true but the OR  model is incorrect.

  If the two models, the OR model and the PS model, are correctly specified, then the two variance forms are equal to 
  \begin{equation*} 
  \V \left\{ \eta(\B{X}, Y, \delta) \right\} = \V \left\{ \E(Y \mid \B{X} ) \right\} + \E\left[ \{ \OP{P}(\delta = 1 \mid \B{X}) \}^{-1} \V \left (Y \mid \B{X} \right ) \right],
  \end{equation*}
  which is equal to the lower bound of the semiparametric efficient estimator considered in \cite{robins1994estimation}. The influence function in (\ref{influence}) can also be used to develop the linearized variance estimation formula for $\hat{\theta}_{\mathrm{APSW}}$ regardless of whether the outcome regression model or the propensity score model holds.

In summary, the influence function in (\ref{influence}) can be reduced to the following limits under each model as follows: 
  \begin{equation}
      \vertarrowboxx{\eta (\bx_i, y_i, \delta_i)}{\text{Correct OR:}}{\text{Correct RP:}}  =  \delta_i  \downvertarrowbox{\underbrace{\omega_i^*}{}}{\pi_{1, i}^{-1}} y_i - \theta  +  (1 - \delta_i \downvertarrowbox{\underbrace{\omega_i^*}{}}{\pi_{1, i}^{-1}}) \bb_i\trans \upvertarrowbox{\overbrace{{\bbeta}^*}{}}{\bbeta} +  (1 - \delta_i /\downvertarrowbox{\underbrace{\pi_{1, i}^{(0)} (\phi^*)}{}}{\pi_{1,i}} ) \bm h_i\trans\upvertarrowbox{\overbrace{\bm \kappa^*}{}}{\bm 0} ,
 \end{equation}
 where $\pi_{1,i} = P(\delta = 1\mid \bx_{i})$,  the upward  arrows  point  to the limits under the correct OR model and the downward arrows  points to the limits under the correct RP model.

  In the next section, we address how to obtain the robust propensity score weighting estimator against model misspecification of the propensity score model and outliers in the outcome regression model.

  \section{Adding outlier-robustness }
  \label{section:RI_info_proj}
  
  In many real cases, outliers exist in addition to missingness. In this case, imposing robustness against outliers is an important practical problem. This section discusses how to allow robust inference against outliers in the outcome variable in the context of doubly robust estimator. In the presence of outliers, one may use the heavy-tailed distribution (e.g. $t$-distribution) \citep{lange1989} to allow robust inference against outliers.  However, it is not straightforward to extend the indirect model calibration condition to the $t$-distribution.

  \cite{basu1998} introduced the  density power divergence as a generalization of Kullback-Leibler divergence to expand the class of statistical models by introducing an additional scale parameter $\gamma$. Density power divergence is also called $\gamma$-power divergence \citep{eguchi2021}. 
  We can use the $\gamma$-power divergence to
  develop an outlier robust imputation method.
  Specifically, we employ the following $M$-estimator  to handle the misspecification of the propensity score model. Define
  \begin{align}\label{objective}
   Q_{\gamma}(\bbeta \mid \blambda ) = \sum_{i=1}^n \delta_i(\hat{d}_i -1 ) \Psi_{\gamma} \left(  y_i - \bb_i^{\T} \bbeta\right)  g_i ( {\blambda}),
   \end{align}
   where $\Psi_{\gamma}$ is an objective function that reduces the effect of outliers, and
   ${g}_i = \exp ( {\bb}_i^{\T} {\blambda} )$ 
   is the adjustment term to achieve the self-efficiency. 
   Thus, under some suitable choice of $\Psi_{\gamma}$, 
   the resulting estimator is not only doubly robust, but also outlier-robust.
  Note that the 
   Huber-type loss function could be used in (\ref{objective}), but it is computationally less attractive as the Huber loss function is non-smooth.

  To incorporate the $\gamma$-divergence, we now consider  the following 
  minimization problem,
  \begin{align}\label{objective new}
   \argmin_{(\bbeta, \sigma^{2})}
   Q_{\gamma}(\bbeta; \sigma^{2} \mid \blambda) =- (2\pi\sigma^{2})^{-  \frac{\gamma}{1+ \gamma} } 
   \sum_{i=1}^n \delta_i ( \hat{d}_i -1) 
  \exp\left\{
  -\frac{\gamma}{  2\sigma^{2}  } (y_{i} - \bb_{i}\trans\bbeta )^{2} 
  \right\} {g}_i (\blambda).
   \end{align}
  That is, in (\ref{objective}), we use 
  \begin{align}\label{Phi}
    \Psi_{\gamma} \left(  x \right)
    = -(2\pi\sigma^{2})^{ - \frac{\gamma}{1+ \gamma} } \exp\left\{
  -\frac{\gamma  }{  2\sigma^{2}  } x^2 
  \right\}.
  \end{align}
  As a result, the optimization problem \eqref{objective new} can be solved by the iterated reweighted least squares method. \textcolor{blue}{That is}, we use 
  \begin{align} 
   \sum_{i=1}^n \delta_i  \left( y_i - \B{b}_i\trans \bbeta \right)( \hat{d}_i-1 ) g_i(  \blambda) q_{\gamma, i} (\bbeta, \sigma^2) \B{b}_i&= \B{0} \label{ceq1} \\ 
    \sum_{i=1}^n \delta_i \left\{  \left( y_i - \B{b}_i\trans \bbeta \right)^2 - \frac{\sigma^{2}}{ 1 + \gamma} \right\}( \hat{d}_i-1 ) g_i(  \blambda) q_{\gamma, i} (\bbeta, \sigma^2) &= 0 \label{ceq2} 
    \end{align} 
  to estimate $\bbeta$ and $\sigma^2$ for given $\blambda$, 
  where
  \begin{equation}
    q_{\gamma,i} (\bbeta, \sigma^2 )  = 
    \exp \{      - 0.5\gamma\sigma^{-2} (y_{i}- \bb_{i}\trans\bbeta )^{2}          \}.
  \label{qi}
  \end{equation}
  
  Note that during the iterative reweighted least squares estimation procedure, if $|y_{i} - \bb_{i}\trans\hat{\bbeta}| \gg 0 $, then the effect of the $i$-th subject for the estimation of $\bbeta$ 
   will be greatly
  mitigated as $\hat{w}_{\gamma, i}$ will be smaller, indicating the robustness of our proposed method. The parameter $\gamma>0$ plays the role of the tuning parameter for robust estimation. As the value of $\gamma$ increases, the estimator becomes more robust but less efficient.

  Let $\hat{\bbeta}_q$ and $\hat{\sigma}_{q}^{2}$ be the solution to the above estimating equations,  (\ref{ceq1}) and (\ref{ceq2}). Further, let $\wh{q}_{\gamma, i}=  {q}_{\gamma, i} (\wh{\bbeta}_q, \wh{\sigma}_{q}^2)$.
  We can construct  robust imputation with $\hat{y}_i=\bb_{i}\trans\hat{\bbeta}_q$ easily. Furthermore, we can use the robust imputation   to construct robust propensity score weights with calibration constraints. Specifically, by (\ref{ceq1}), we can obtain 
  \begin{equation}
   \sum_{i=1}^n \delta_i\left ( y_i -\B{\OP{b}}_i^{\T} \wh{\bbeta}_q \right ) \B{\OP{b}}_i (\wh{d}_i -1 )  g_i( \wh{\B{\lambda}} )\wh{q}_{\gamma, i}
  = \B{0}.
  \label{cond9}
  \end{equation}
In a similar fashion of derivation for   Lemma~\ref{lem2}, it yields that
  \begin{equation}
     \sum_{i=1}^{n} \left\{ \delta_i y_i + (1-\delta_i) \hat{y}_{i} \right\}
     = \sum_{i=1}^n \delta_i \hat{\omega}_{\gamma, i} y_i     , 
    \label{eq43}
  \end{equation}
  where $\hat{y}_i = \B{b}_i\trans \hat{\bbeta}_q$ 
  and 
  \begin{equation}
   {\omega}_{\gamma, i} (\bx_{i}; \blambda, \bbeta, \sigma^2,  \wh{\bphi}) = 1+ ( d_i(\wh{\bphi}) -1 ) {g}_i (\blambda)  {q}_{\gamma, i} (\bbeta, \sigma^2) .
   \label{wgt3}
   \end{equation}
   Note that (\ref{eq43}) is the self-efficiency of the propensity score weights in (\ref{wgt3}). 
 The final PS weights satisfy 
  \begin{equation}
  \sum_{i=1}^n \delta_i \wh{\omega}_{\gamma, i} \B{b}_i = \sum_{i=1}^n \B{b}_i.
    \label{calib5}
  \end{equation}
  Thus, for given $\wh{q}_{\gamma, i}$'s, we can compute $\wh{\blambda}$ from calibration constraints in (\ref{calib5}). That is, we can treat (\ref{ceq1}), (\ref{ceq2}), and (\ref{calib5}) as the estimating equations for $\bbeta$, $\sigma^2$ and $\blambda$. Note that (\ref{wgt3}) is equivalent to (\ref{wgt2}) for $q_{\gamma, i}=1$. The additional factor ${q}_{\gamma, i} (\bbeta, \sigma^2)$ controls the effect of outliers in the final weighting.


  Let $\bbeta^{\ast}$ and $\sigma^{\ast 2}$ be the probability limits of $\wh{\bbeta}_q$ and $\wh{\sigma}_{q}^{2}$,
  respectively. 
  The  following theorem presents the Taylor linearization for our proposed estimator in \eqref{eq43}.
  \begin{theorem}\label{TR} 
    Let $\wh{\theta}_{\mathrm{APSW}, \gamma}=n^{-1} \sum_{i=1}^n \delta_i \hat{\omega}_{\gamma, i} y_i$  be the propensity score weighting  estimator under the augmented propensity score model in 
    (\ref{augps}) with $\wh{\bphi}$ and $\wh{\blambda}$ satisfying (\ref{ee5}) and (\ref{cond9}), 
    $\wh{\bbeta}_q$ and $\wh{\sigma}_q^{2}$ minimizing \eqref{objective new}  respectively. 
    Under the regularity conditions described in the Appendix, we have
    \begin{align}
     \wh{\theta}_{\rm APSW, \gamma} - \theta 
     = 
     \frac{1}{n} \sum_{i=1}^n  {\eta}_{\gamma} (\bx_i, y_i, \delta_i) + o_p(n^{-1/2}), \label{robust} 
    \end{align}
    where 
\begin{align}
       {\eta}_{\gamma} (\bx_i, y_i, \delta_i) 	 &= \bb_{i}\trans \bmu - \theta 
      + \delta_{i} \omega_{\gamma, i}^{*}
      (y_{i} - \bb_{i}\trans \bmu^{\ast})
        +   \left\{
      1 - \delta_{i} d_{i}(\bphi^{\ast})  
      \right\}\bkappa^{\ast {\rm T}}\bh(\bx_{i}; \bphi^{\ast}) \notag \\        &\quad+ 
        \delta_{i}   \{\omega_{\gamma, i}^{*} - 1\}(y_{i} - \bb_{i}\trans\bbeta^{\ast})
      \bb_{i}\trans\bzeta^{\ast}\notag\\
      &\quad +  \delta_{i}\{\omega_{\gamma, i}^{*} - 1\}(y_{i} - \bb_{i}\trans \bbeta^{\ast})^{2} 
      \nu^{\ast}\left\{
        (y_{i} - \bb_{i}\trans \bbeta^{\ast})^{2} - \sigma^{\ast 2}/(1  + \gamma)
      \right\} ,
      \label{influence robust}
      \end{align}
 where $\omega_{\gamma, i}^{*} = \omega_{\gamma, i}(\bx_{i}; \blambda^{\ast},
      \bbeta^{\ast}, \sigma^{\ast 2}, \bphi^{\ast})$ and $\bmu^{\ast}$, $\bkappa^{\ast}$, $\nu^{\ast}$, $\bzeta^{\star}$ are the probability limits 
    of the solutions for the equations presented in the Appendix. 
    \label{theoremrobust}
    \end{theorem}

  Note that the first line and the last line has the same structure as in \eqref{influence} of Theorem~\ref{T1}. 
  The other two lines are the additional part of the influence functions that reflect the effect of the gamma-divergence model. Specifically, 
  the second line reflects the estimation effect of $\wh{\bbeta}_q$, and
  the third line reflects the estimation effect of $\wh{\sigma}_q^{2}$. 

  Regarding the choice of the tuning parameter $\gamma$, we can use cross-validation to choose the optimal $\gamma$. Specifically, we can randomly partition the sample  into $K$-groups ($S_1, \cdots, S_K$) and then compute 
  $$ 
  \mbox{MSPE}(\gamma) = \sum_{k=1}^K \sum_{i \in S_k} \delta_i \left\{ y_i - \hat{y}_i^{(-k)} (\gamma) \right\}^2  
  $$
  where $\hat{y}_i^{(-k)} (\gamma)= 
  {\bb}_i^{\T} \wh{\bbeta}_q^{(-k)}$ and $\hat{\bbeta}_q^{(-k)}$ is obtained by solving the same estimation formula using the sample in $S_k^c$. 
  The minimizer of $\mbox{MSPE}(\gamma)$ can be used as the optimal value of the tuning parameter. 
  Furthermore, instead of cross-validation, we can use other approximation-based methods \citep{sugasawa2021, Basak2020} that are computationally less expensive. 
  As we can find in our simulation study in Section~\ref{section:RI_simulation}, the optimal choice of $\gamma$ is not critical. The simulation result shows similar performances for different values of $\gamma$.  
  
  For variance estimation, we can use the influence function in \eqref{influence robust} to construct a linearization variance estimator.
  Specifically, the linearization variance estimator for $\hat{\theta}_{\rm APSW, \gamma}$ for a given $\gamma$ is 
  \begin{align}
    \widehat{\rm V} (\wh{\theta}_{\rm{APSW}, \gamma})
    = \frac{1}{n(n-1)}\sum_{i=1}^{n} (\wh{\eta}_{\gamma}(\bx_{i}, y_{i}, \delta_{i}) - \bar{\eta}_{\gamma})^{2}, \label{linearized variance estimation}
  \end{align}
  where
  \begin{align}
      &\quad  \wh{\eta}_{\gamma} (\bx_i, y_i, \delta_i) \notag\\ 
      &= \bb_{i}\trans \wh{\bmu}  
      + \delta_{i} \hat{\omega}_{\gamma, i}
      (y_{i} - \bb_{i}\trans \wh{\bmu})
        +   \left\{
      1 - \delta_{i} d_{i}(\wh{\bphi})  
      \right\}\wh{\bkappa}^{ {\rm T}}\bh(\bx_{i}; \wh{\bphi}) \notag \\        &\quad+ 
        \delta_{i}   \{\hat{\omega}_{\gamma, i} - 1\}(y_{i} - \bb_{i}\trans\wh{\bbeta}_{q})
      \bb_{i}\trans\wh{\bzeta}\notag\\
      &\quad +  \delta_{i}\{\hat{\omega}_{\gamma, i} - 1\}(y_{i} - \bb_{i}\trans \wh{\bbeta}_{q})^{2} 
      \wh{\nu}\left\{
        (y_{i} - \bb_{i}\trans \wh{\bbeta}_{q})^{2} - \wh{\sigma}_{q}^{ 2}/(1  + \gamma)
      \right\},
      \label{estimated influence robust}
  \end{align}
 $\hat{\omega}_{\gamma, i} = \omega_{\gamma, i}(\bx_{i}; \wh{\blambda},
      \wh{\bbeta}_{q}, \wh{\sigma}_{q}^{ 2}, \wh{\bphi})$
 and 
  \begin{align}
    \bar{\eta}_{\gamma} = \frac{1}{n}
    \sum_{i=1}^{n}   \wh{\eta}_{\gamma} (\bx_i, y_i, \delta_i).\notag
  \end{align}
  Note that $\wh{\bbeta}_{q}$, $\wh{\sigma}_{q}^{2}$ and $\wh{\blambda}$ are estimated from \eqref{ceq1},  \eqref{ceq2} and \eqref{calib5}, $\wh{\bmu}$, $\wh{\bkappa}$, $\wh{\nu}$ and  $\wh{\bzeta}$ can be estimated from the procedure listed in the Appendix. A flowchart of  the inference procedure for the robust augmented PSW  estimator is presented in Fig~\ref{flowchart APS gamma}, where the equations (E.2) and (E.6) are estimating equations presented in Appendix~\ref{appendix:E}.

 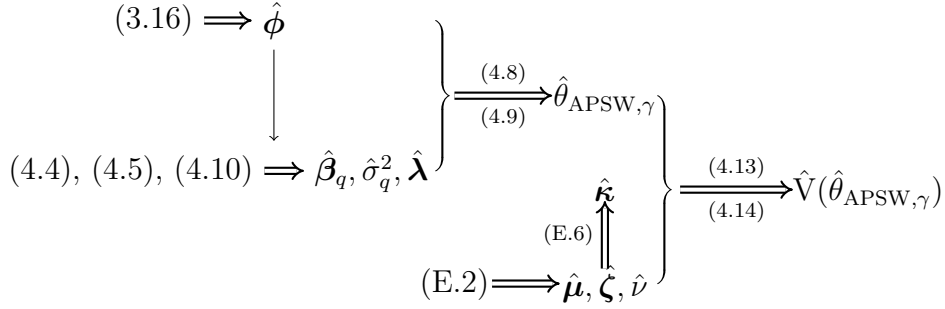
\begin{figure}
 \begin{center}
 \begin{tikzpicture}[node distance=2cm]
 
 \node (working)   {(3.16)};
 \node (phi) at (1.6,0) {$\hat{\bphi}$};
 
 \node (original) at (-0.3,-2) {(4.4), (4.5), (4.10)};

 \node (lambda) at (2.9, -2) {$\hat{\bbeta}_{q}, \hat{\sigma}^{2}_{q}, \hat{\blambda}$};

 \node (APS) at (6, -1) {$\hat{\theta}_{\rm{APSW}, \gamma}$};
 
 \node (variance) at (9.5, -2.25) {$\wh{\rm{V}}(\hat{\theta}_{\rm{APSW}, \gamma})$};
 
 \node (E.2) at (4, -3.5) { (E.2) };
 \node (mu) at (6, -3.5) { $\hat{\bmu}, \hat{\bzeta}, \hat{\nu}$ };
 \node (kappa) at (6, -2.25) { $\hat{\bkappa}$ };


 \draw [-implies,double equal sign distance, thick] (working) -- (phi);
 
 \draw [-implies,double equal sign distance, thick] (original) --  (lambda);
 
 \draw [->] (phi) -- (1.6, -1.6);
 
 \draw [decorate, thick,
     decoration = {calligraphic brace, 
         raise=10pt,
         amplitude=5pt}] (3.4, 0) --  (3.4,-2);

 \draw [-implies,double equal sign distance, thick](4,-1) -- (5.3,-1) node[midway, above]{\scriptsize{(4.8)}} node[midway, below]{\scriptsize{(4.9)}};

 \draw [-implies,double equal sign distance, thick](6,-3.3) -- (6,-2.4) node[midway, left]{\scriptsize{(E.6)}};
 
 \draw [-implies,double equal sign distance, thick](4.5,-3.5) -- (5.4,-3.5);

 \draw [decorate, thick,
     decoration = {calligraphic brace, 
         raise=20pt,
         amplitude=5pt}] (6,-1) --  (6,-3.5);

 \draw [-implies,double equal sign distance, thick](7,-2.25) -- (8.5,-2.25) node[midway, above]{\scriptsize{(4.13)}} node[midway, below]{\scriptsize{(4.14)}};
 
 \end{tikzpicture}
  \caption{Illustration of the inference procedure for the 
 robust augmented PSW  estimator.}
 \label{flowchart APS gamma}
 \end{center}
 \end{figure}

\section{Empirical studies}
\label{section:RI_simulation}
\subsection{Simulation study}
\label{section:robust_model}

We examine the performance of proposed methods under various propensity score models. Let ${\bX} = (X_{1}, X_{2}, X_{3}, X_{4}, X_{5})\trans$, where  $\bX \sim N(\bzero, \bI_{5\times 5})$ follow the standard multivariate normal distribution.  The simulation experiment can be described as a $2 \times 2$ factorial design with two factors.  The
sample size is $100$ and the Monte Carlo sample size is $1,000$. The first factor is the outcome regression model (OM1, OM2) and the second factor is the propensity score model (PM1, PM2).

The models for generating $Y$ are described as follows: OM1, where $Y \mid {\bX}$ follows a normal distribution with mean
$
    \E(Y \mid  {\bX}) = 1 + X_{1} + X_{2} + X_{3} + X_{4} + X_{5} 
$
and variance 1;
OM2, where $Y \mid {\bX} $ follows a normal distribution with
$
    \E(Y \mid {\bX} ) = 1 + 0.25 \{\cos(3\pi X_1) + 1\}X_{2}^{3} + 0.25 \{\sin(3\pi X_4) - 1\}X_{3}^{3} + \cos(3\pi X_{5})
$
and variance 1. 
For calibration, we use ${\bb}(\bX)=(1, X_1, X_2, X_3, X_4, X_5)\trans$ as the calibration variable. Thus, the implicit model calibration using ${\bb}(\bX)=(1, X_1, X_2, X_3, X_4, X_5)\trans$ is justified under the OM1 outcome model, but it is incorrectly specified under the OM2 outcome model. 
In addition, the models for generating $\delta$ can be described as follows: 
PM1, where  $\delta \mid  {\bX}$ follows Bernoulli distribution with
$
\mathrm{logit}\{P(\delta = 1 \mid {\bX})\}=  \phi_0 + 0.25 X_{1} + 0.25X_{2}+ 0.25 X_{3} + 0.25X_{4}+ 0.25 X_{5},
$
where $\phi_0$ is chosen to achieve 60\% response rates;
PM2, where $\delta \mid {\bX}$ follows the Bernoulli distribution with
$
P( \delta=1 \mid {\bX} ) = 
0.8$  if  $a + X_1 + X_2 > 0$, and 
$
P( \delta=1 \mid {\bX} ) = 0.4$ otherwise, 
where $a$ is chosen to achieve 60\% response rates. 
Further, we use 
\begin{equation}
    {\mathrm{logit}} \left \{ \pi_{1}^{(0)}({\bX}; {\bphi}) \right \} = \phi_{0} + \phi_{1} X_{1} + \phi_{2} X_{2}+ \phi_{3} X_{3} + \phi_{4} X_{4}+ \phi_{5} X_{5}.
    \label{wps}
\end{equation} 
as the working model for the propensity score function, where $\mathrm{logit}(x)= \log\{x/(1-x)\}$. Thus,  the working propensity score model is correctly specified under PM1 only.

For each sample, we compute the following propensity score estimators. 
\begin{enumerate}[label=\textnormal{(\roman*)}]
    \item \textit{Mean of the complete cases} (CC): $\hat{\theta}_{\mathrm{CC}} = n_{1}^{-1}\sum_{i=1}^{n}\delta_{i}Y_{i}$, that is, mean of observed $Y_{i}$'s, where $n_{1} = \sum_{i=1}^{n}\delta_{i}$. 
        \item \textit{Maximum likelihood estimation} (MLE): the classical propensity score estimator 
         $\hat{\theta}_{\mathrm{PSW}} = n^{-1} \sum_{i=1}^{n} \delta_{i} \hat{d}_{i} Y_{i}$  using 
          $\hat{d}_{i}= 1/\pi_{1}^{(0)}({\bX}_{i}; \hat{{\bphi}})$ as the estimated propensity score weight, where $\hat{{\bphi}}$ is the maximum likelihood estimate of ${\bphi}$ from the working propensity score model in (\ref{wps}).
    \item \textit{Entropy balancing method in \cite{hainmueller2012}}: 
    $\hat{\theta}_{\mathrm{HM}} = n^{-1}\sum_{i=1}^{n}\delta_{i}\hat{\omega}_{i}Y_{i}$, where $\hat{\omega}_{i}$ is obtained from \eqref{entropy}.
    \item  \sloppy \textit{Regularized calibration method in \cite{tan2020}}: $\hat{\theta}_{\mathrm{Tan}} = n^{-1} \sum_{i=1}^{n} \delta_{i} \{\pi_{1}^{(0)}({\bX}_{i}; \hat{{\bphi}})\}^{-1} Y_{i}$, where the logistic regression coefficients $\hat{\bphi}$ are obtained by minimizing  $\ell_{\mathrm{CAL}}(\bphi) = 
        \sum_{i=1}^{n} [  \delta_{i}\exp\{ -\bphi^{\T} f(\bX_{i}) \}
        - (1-\delta_{i})\bphi\trans f(\bX_{i}) ]$, where we take the identity mapping for $f(\cdot)$ in this simulation.  
    \item \textit{Augmented propensity score weighting estimator} (APS):   
    $\hat{\theta}_{\mathrm{APSW}} = n^{-1} \sum_{i=1}^{n} \delta_{i} \hat{\omega}_{i}^{\ast} Y_{i}$ 
    using $\hat{\omega}_i^{\ast}$ in (\ref{wgt2}).
        \item   \textit{Augmented propensity score weighting estimator with $\gamma$-divergence} ($\mathrm{APS}_{\gamma}$): the augmented propensity score weighting estimator in \eqref{eq43} using $\gamma = 0.3, 0.5, 0.7, 1$. 
    \end{enumerate}

    \begin{figure}
      \centering
      \begin{subfigure}{.4\linewidth}
        \caption{OM1PM1}
        \includegraphics[scale=0.35]{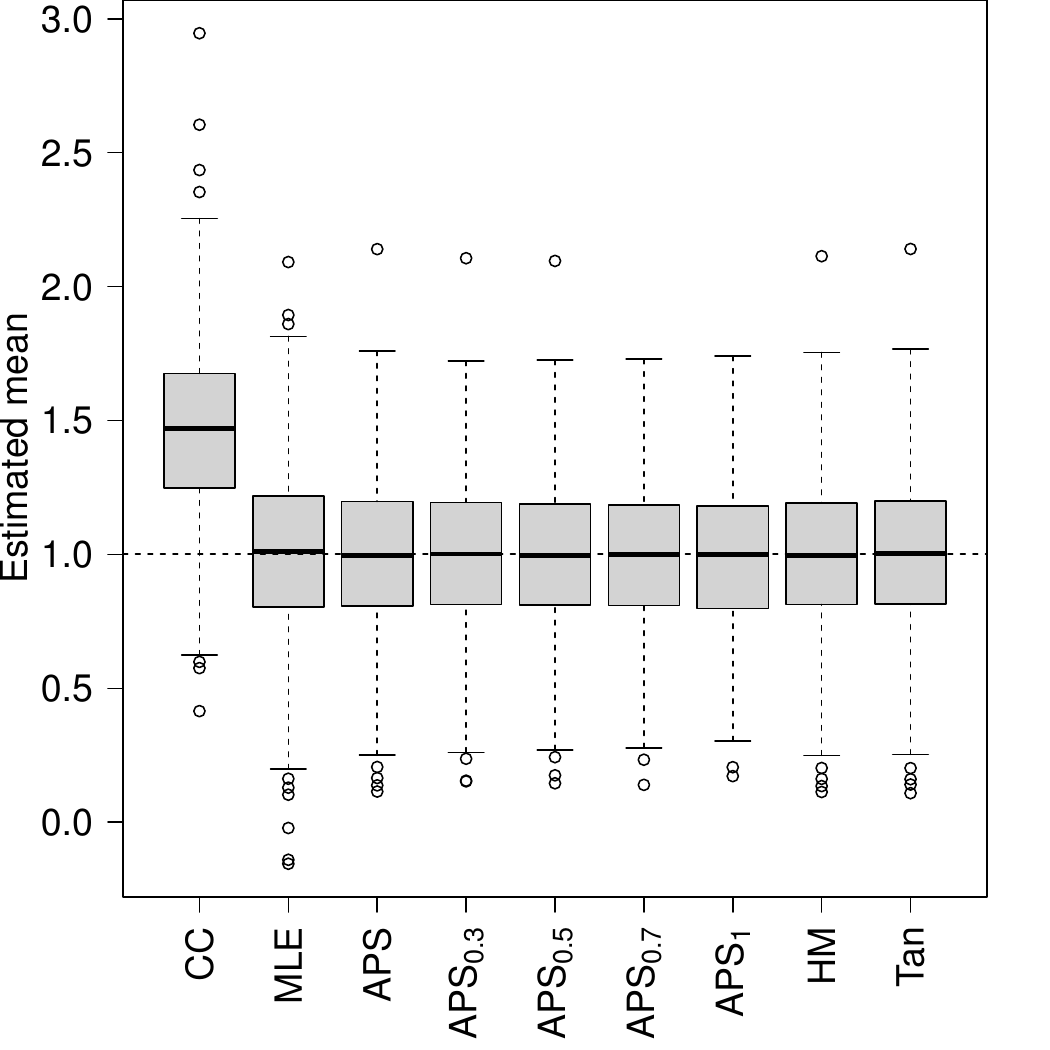}
      \end{subfigure}
      \hskip4em
      \begin{subfigure}{.4\linewidth}
        \caption{OM2PM1}
        \includegraphics[scale=0.35]{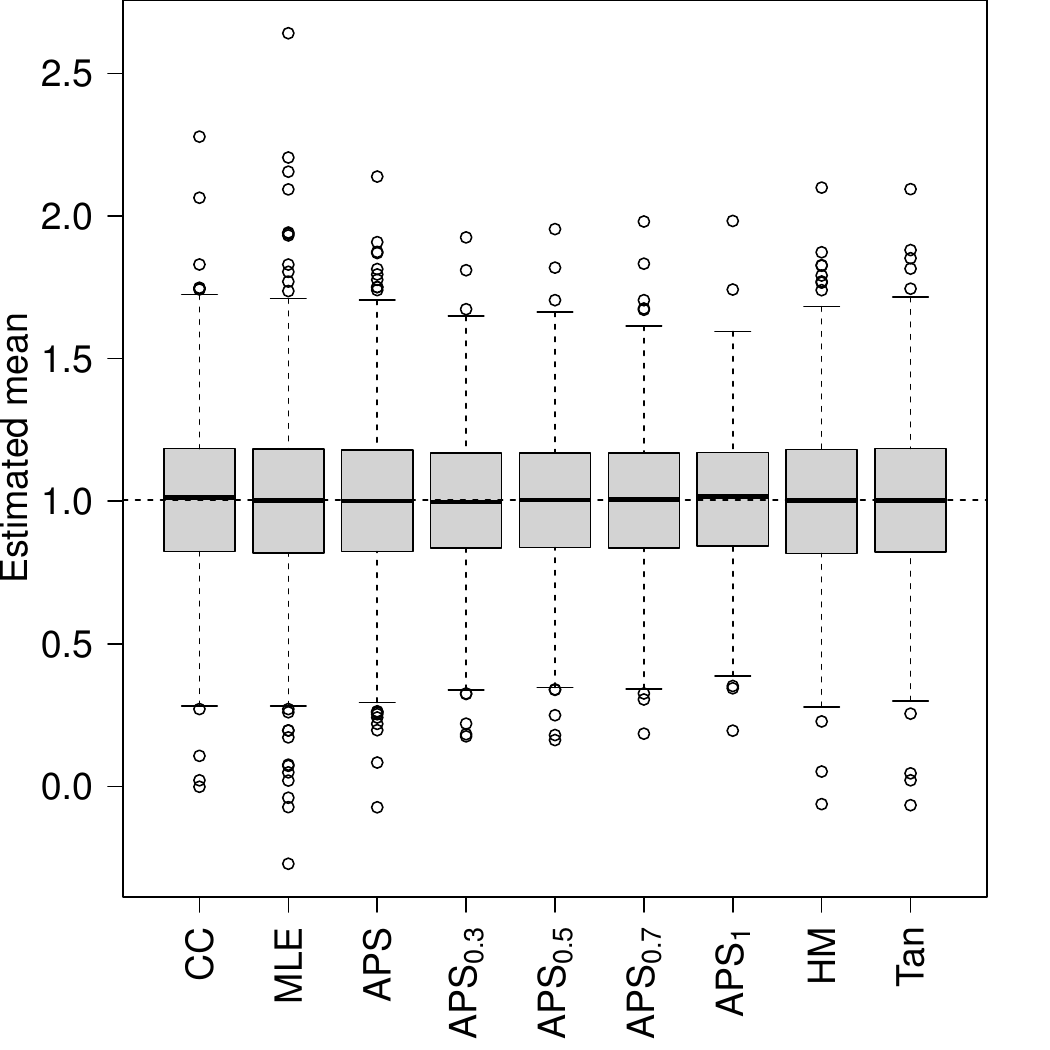}
      \end{subfigure}\\
      \begin{subfigure}{.4\linewidth}
        \caption{OM1PM2}
        \includegraphics[scale=0.35]{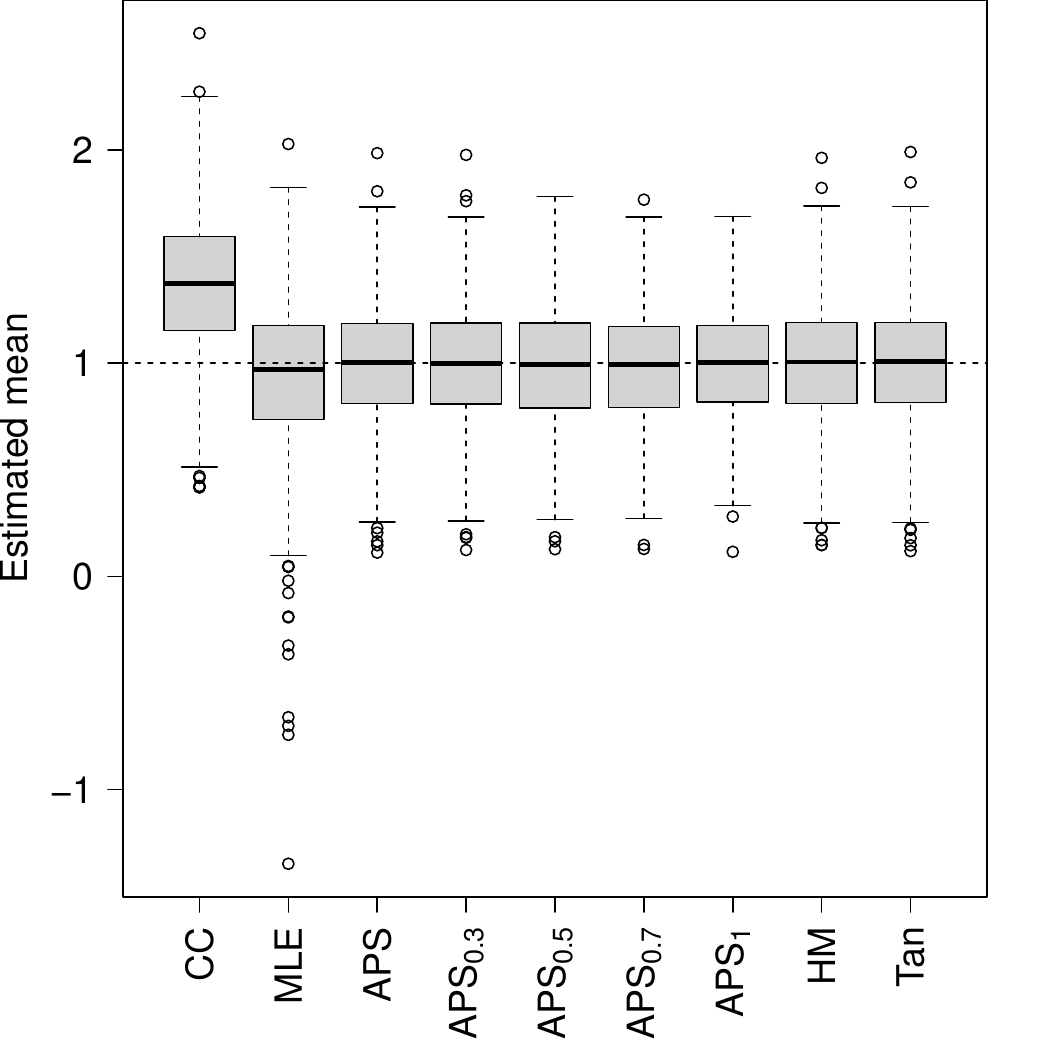}
      \end{subfigure}
      \hskip4em
      \begin{subfigure}{.4\linewidth}
        \caption{OM2PM2}
        \includegraphics[scale=0.35]{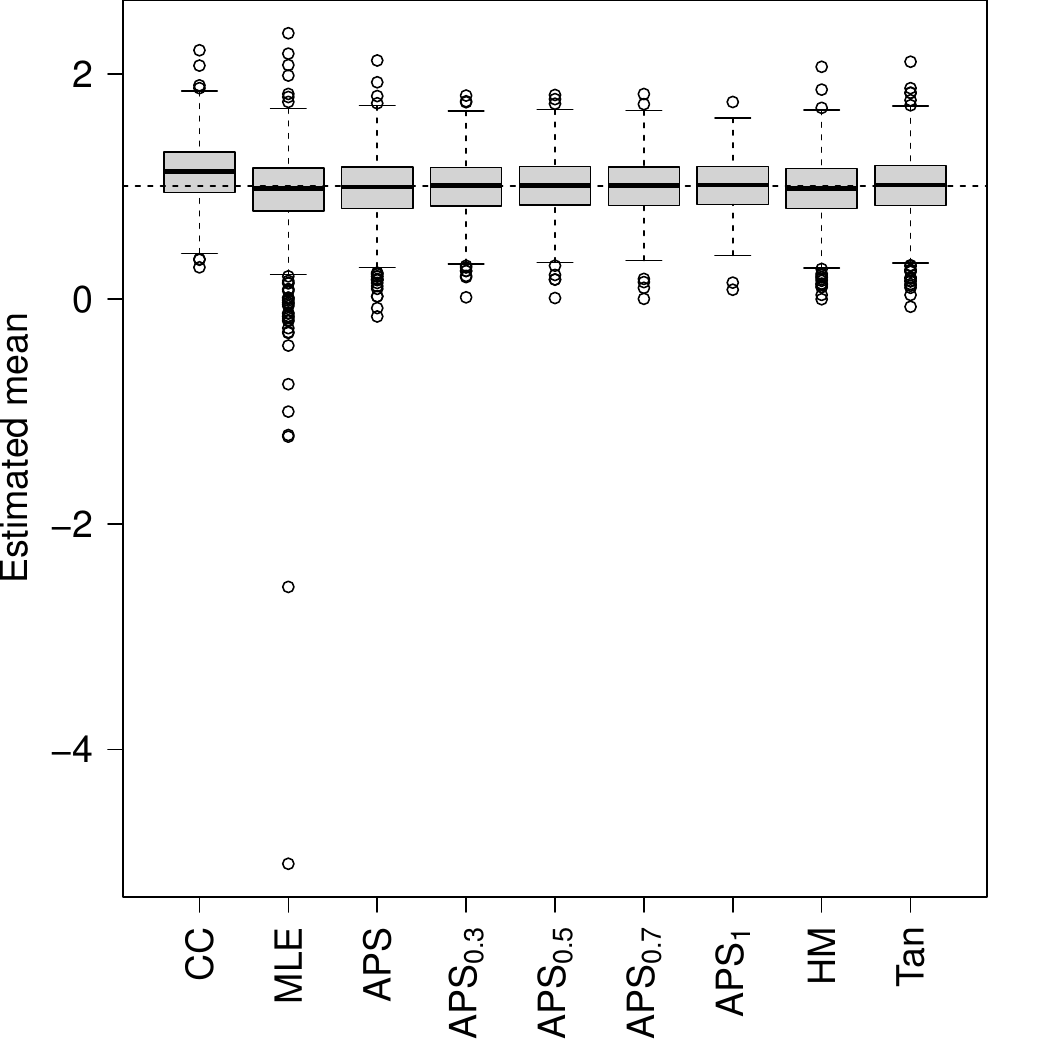}
      \end{subfigure}
      \caption{Boxplots for estimators comparison with sample size $100$ and Monte Carlo sample $1,000$. The errors from outcome models are generated from \textit{i.i.d.} standard Gaussian distribution. CC: mean of the complete cases; MLE: maximum likelihood estimation; APS: augmented propensity score weighting estimator; $\mathrm{APS}_{\gamma}$: augmented propensity score weighting with  $\gamma$-divergence estimator, for  $\gamma = 0.3, 0.5, 0.7, 1$; HM: entropy balancing method in Hainmueller (2012); Tan: regularized calibration method in Tan (2019).}
      \label{fig:clear}
    \end{figure}

\begin{table}[ht]
  \centering
    \caption{Bias, standard error (S.E.), root mean square error (RMSE)  for estimators comparison with sample size $100$ and Monte Carlo sample $1,000$. All criteria are multiplied by 10. The errors from outcome models are generated from \textit{i.i.d.} standard Gaussian distribution.  CC: mean of the complete cases; MLE: maximum likelihood estimation; APS: augmented propensity score weighting estimator; $\mathrm{APS}_{\gamma}$: augmented propensity score weighting with $\gamma$-divergence estimator, for  $\gamma = 0.3, 0.5, 0.7, 1$; HM: entropy balancing method in Hainmueller (2012); Tan: regularized calibration method in Tan (2019) }
    \label{tab:clear}
    \begin{tabular}{ccrrrrrrrrr}
      \hline
      &  & \multicolumn{9}{c}{Method}\\
      \cline{3-11}
      \multirow{-2}{*}{Model}  & \multirow{-2}{*}{Criteria} & CC & MLE & APS & ${\rm APS}_{0.3}$ & ${\rm APS}_{0.5}$ & ${\rm APS}_{0.7}$ & ${\rm APS}_{1}$ & HM & Tan \\ 
      \hline
  &Bias & 4.62 & 0.04 & -0.01 & -0.00 & -0.02 & -0.02 & -0.00 & -0.01 & 0.03 \\ 
  &S.E. & 3.26 & 3.02 & 2.77 & 2.72 & 2.72 & 2.75 & 2.72 & 2.76 & 2.77 \\ 
  \multirow{-3}{*}{OM1PM1}&RMSE& 5.65 & 3.02 & 2.77 & 2.72 & 2.71 & 2.74 & 2.72 & 2.76 & 2.77 \\ 
  \cline{2-11}
  &Bias& 0.03 & -0.02 & -0.03 & -0.02 & -0.01 & 0.01 & -0.03 & -0.03 & -0.02 \\ 
  &S.E. & 2.75 & 2.95 & 2.78 & 2.48 & 2.48 & 2.47 & 2.45 & 2.70 & 2.69 \\ 
  \multirow{-3}{*}{OM2PM1}&RMSE& 2.75 & 2.95 & 2.78 & 2.48 & 2.47 & 2.47 & 2.45 & 2.70 & 2.69 \\ 
  \cline{2-11}
  &Bias& 3.72 & -0.54 & -0.02 & -0.01 & -0.05 & -0.05 & -0.02 & -0.03 & 0.01 \\ 
  &S.E. & 3.28 & 3.51 & 2.76 & 2.76 & 2.76 & 2.71 & 2.73 & 2.75 & 2.75 \\ 
  \multirow{-3}{*}{OM1PM2}&RMSE& 4.96 & 3.55 & 2.76 & 2.76 & 2.76 & 2.72 & 2.73 & 2.75 & 2.75 \\ 
  \cline{2-11}
  &Bias& 1.24 & -0.61 & -0.22 & -0.08 & -0.03 & -0.04 & -0.04 & -0.26 & 0.01 \\ 
  &S.E. & 2.76 & 4.19 & 2.93 & 2.60 & 2.58 & 2.59 & 2.55 & 2.81 & 2.83 \\ 
  \multirow{-3}{*}{OM2PM2}&RMSE& 3.03 & 4.23 & 2.94 & 2.60 & 2.58 & 2.59 & 2.54 & 2.82 & 2.83 \\ 
     \hline
  \end{tabular}
  \end{table}

    \begin{figure}
      \centering
      \begin{subfigure}{.4\linewidth}
        \caption{OM1PM1}
        \includegraphics[scale=0.35]{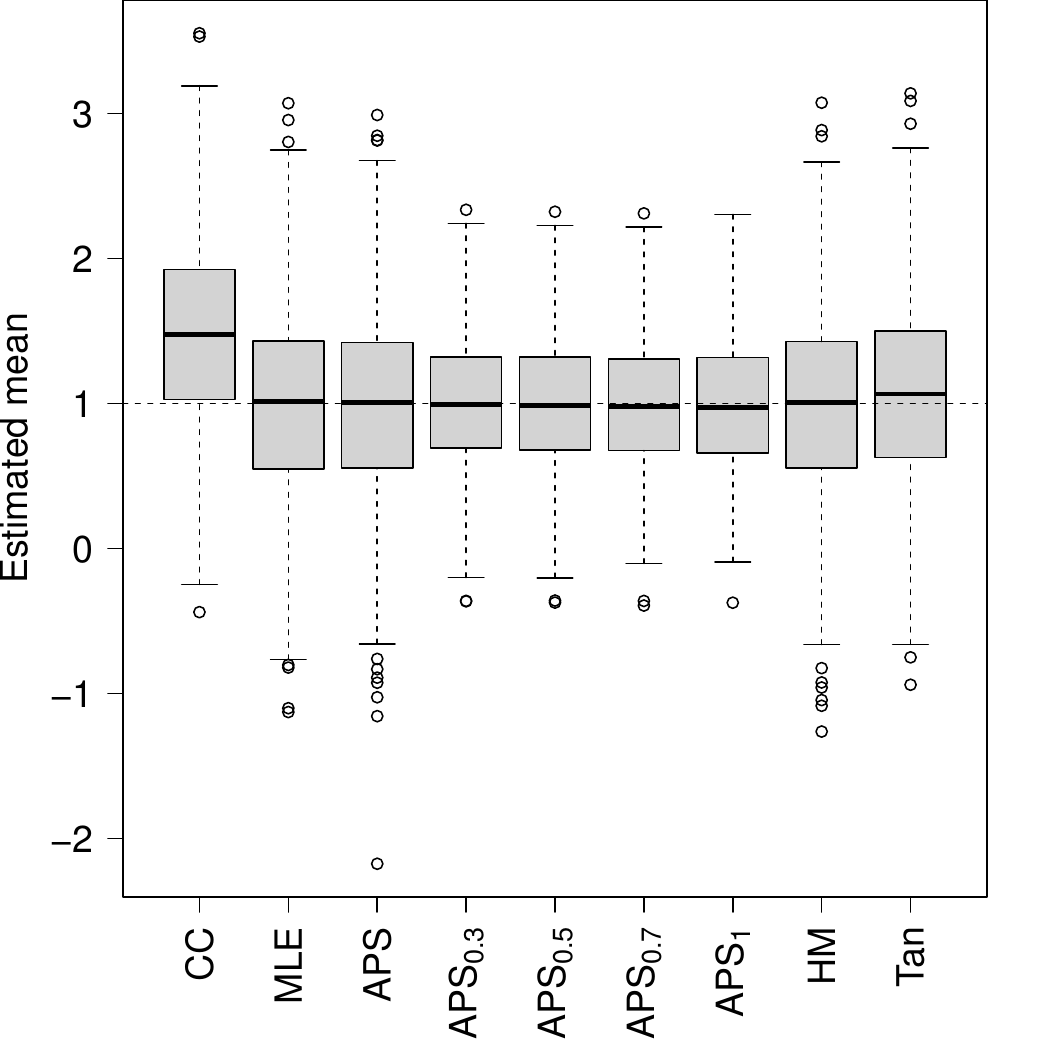}
      \end{subfigure}
      \hskip4em
      \begin{subfigure}{.4\linewidth}
        \caption{OM2PM1}
        \includegraphics[scale=0.35]{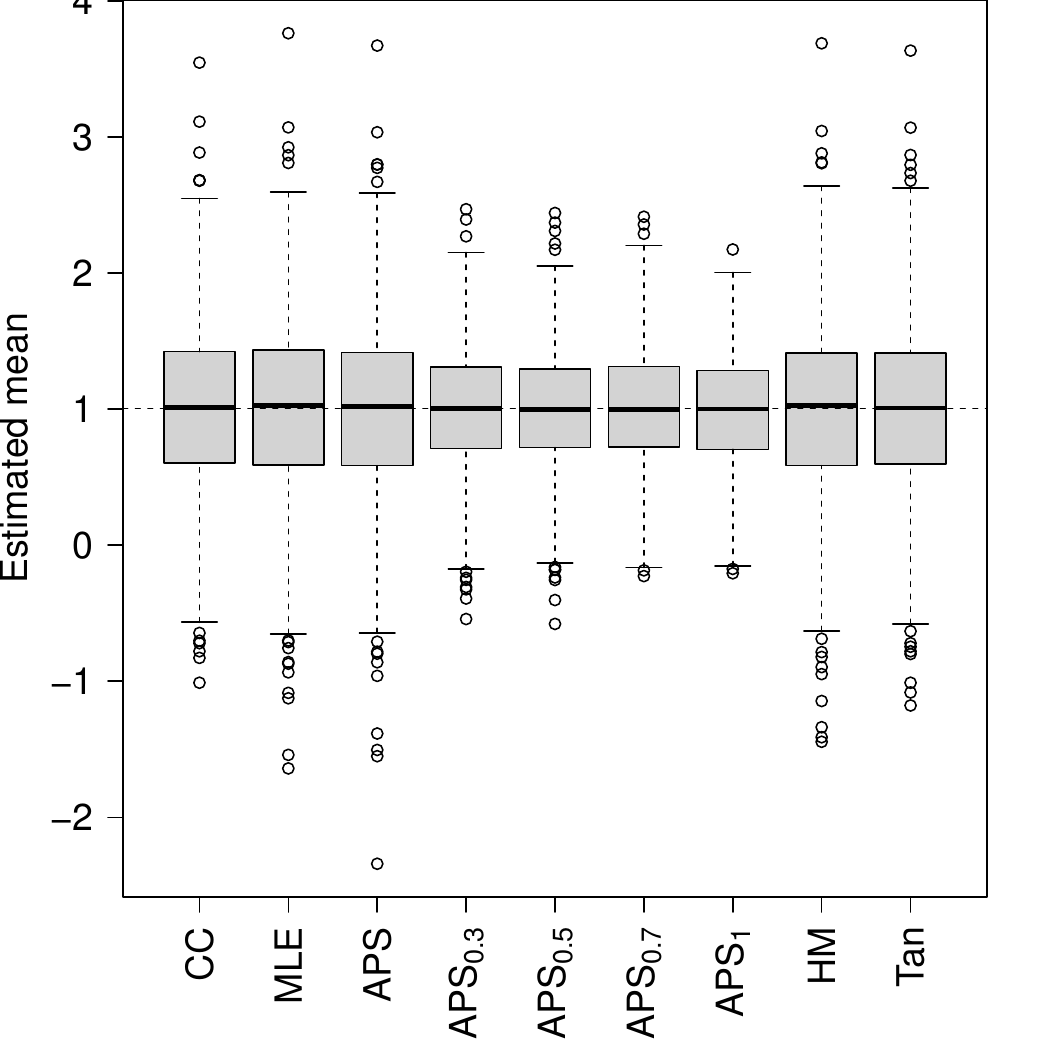}
      \end{subfigure}\\
      \begin{subfigure}{.4\linewidth}
        \caption{OM1PM2}
        \includegraphics[scale=0.35]{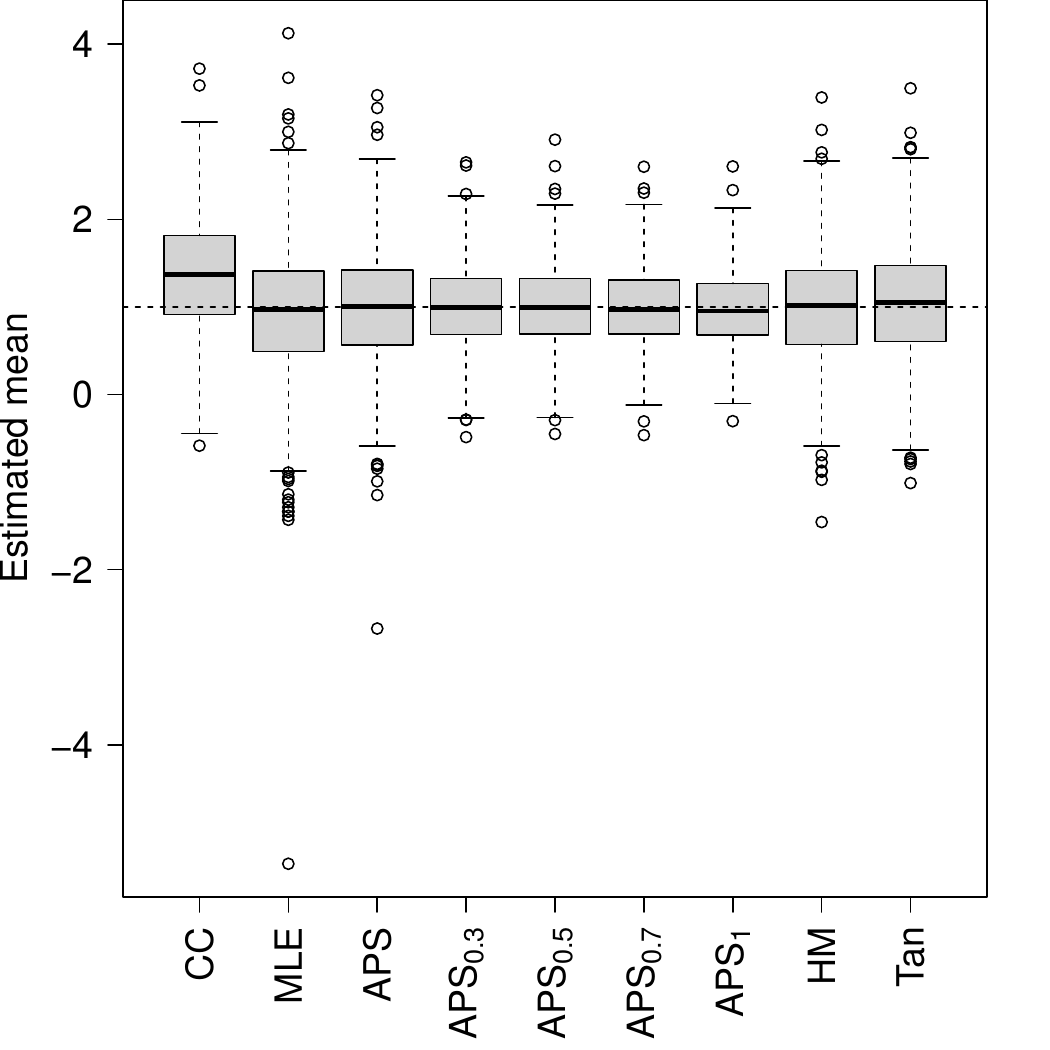}
      \end{subfigure}
      \hskip4em
      \begin{subfigure}{.4\linewidth}
        \caption{OM2PM2}
        \includegraphics[scale=0.35]{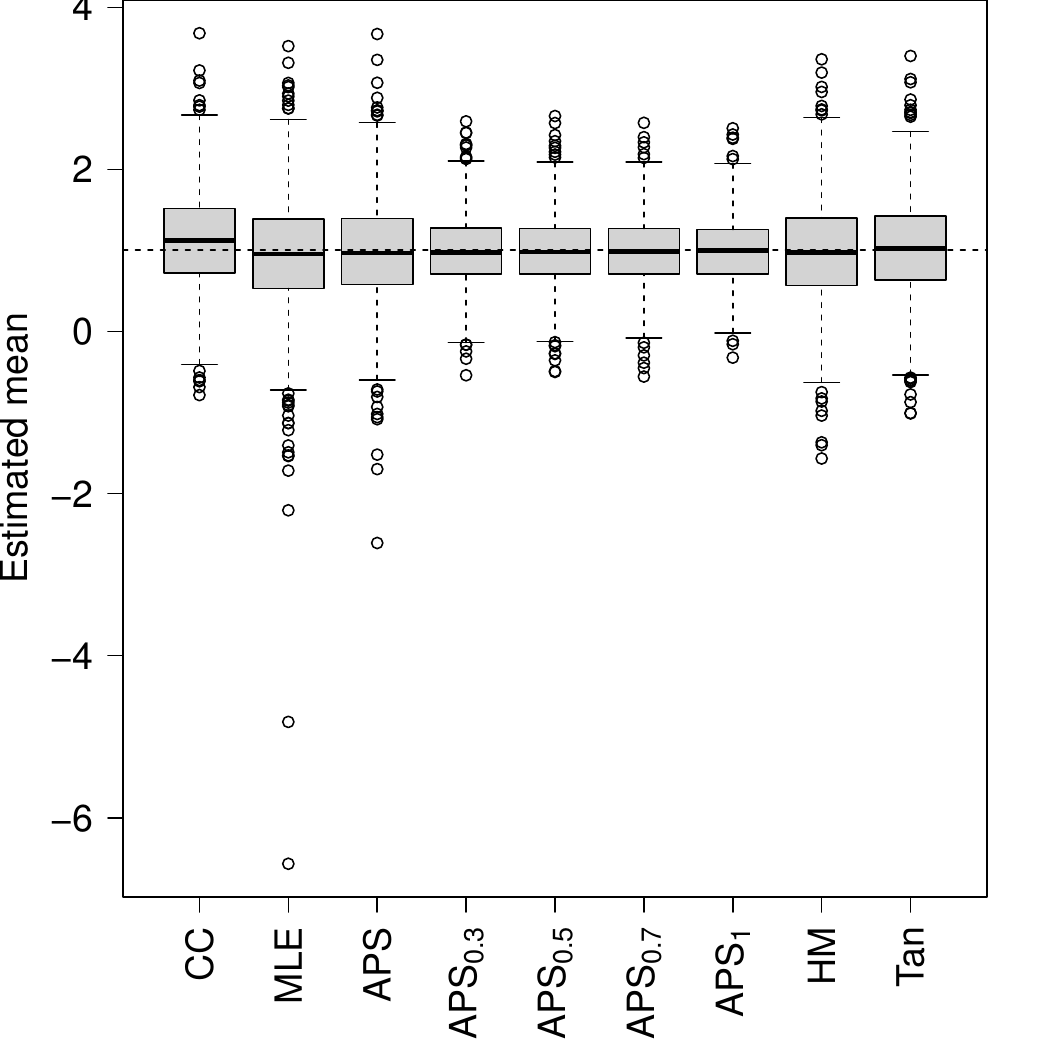}
      \end{subfigure}
      \caption{Boxplots for estimators comparison with sample size $100$ and Monte Carlo sample $1,000$; 20\% of the samples are contaminated with the  additive  noise  that eight times a random variable generated  from Student's t distribution with degree freedom of 8. The errors from outcome models are generated from \textit{i.i.d.} standard Gaussian distribution. CC: mean of the complete cases; MLE: maximum likelihood estimation; APS: augmented propensity score weighting estimator; $\mathrm{APS}_{\gamma}$: augmented propensity score weighting with $\gamma$-divergence estimator, for  $\gamma = 0.3, 0.5, 0.7, 1$; HM: entropy balancing method in Hainmueller (2012); Tan: regularized calibration method in Tan (2019).}
      \label{fig:contaminated}
    \end{figure}

\begin{table}[ht]
		\centering
		\caption{Bias, standard error (S.E.), root mean square error (RMSE)  for estimators comparison with sample size $100$ and Monte Carlo sample $1,000$; 20\% of the samples are contaminated with the  additive noise that eight times a random variable generated  from Student's t distribution with degree freedom of 8. The errors from outcome models are generated from \textit{i.i.d.} standard Gaussian distribution.  All criteria are multiplied by 10. CC: mean of the complete cases;  MLE: maximum likelihood estimation; APS: augmented propensity score weighting estimator; $\mathrm{APS}_{\gamma}$: augmented propensity score weighting with $\gamma$-divergence estimator, for  $\gamma = 0.3, 0.5, 0.7, 1$; HM: entropy balancing method in Hainmueller (2012); Tan: regularized calibration method in Tan (2019)}
		\label{tab:contaminated}
		\begin{tabular}{ccrrrrrrrrr}
			\hline
			&  & \multicolumn{9}{c}{Method}\\
		\cline{3-11}
			\multirow{-2}{*}{Model} & \multirow{-2}{*}{Criteria} & CC & MLE & APS & ${\rm APS}_{0.3}$ & ${\rm APS}_{0.5}$ & ${\rm APS}_{0.7}$ & ${\rm APS}_{1}$ & HM & Tan \\ 
			\hline
  & Bias& 4.62 & 0.00 & -0.05 & -0.10 & -0.09 & -0.14 & -0.13 & -0.04 & 0.69 \\ 
  & S.E. & 6.48 & 6.62 & 6.59 & 4.52 & 4.46 & 4.44 & 4.36 & 6.51 & 6.50 \\ 
 \multirow{-3}{*}{OM1PM1} & RMSE& 7.95 & 6.62 & 6.59 & 4.52 & 4.46 & 4.44 & 4.36 & 6.51 & 6.53 \\ 
 \cline{2-11}
  & Bias& 0.03 & -0.06 & -0.06 & -0.01 & -0.07 & -0.01 & -0.18 & -0.06 & -0.05 \\ 
  & S.E. & 6.11 & 6.51 & 6.54 & 4.45 & 4.38 & 4.37 & 4.21 & 6.43 & 6.24 \\ 
\multirow{-3}{*}{OM2PM1}  & RMSE& 6.11 & 6.51 & 6.53 & 4.45 & 4.37 & 4.37 & 4.21 & 6.42 & 6.23 \\ 
\cline{2-11}
  & Bias& 3.66 & -0.64 & -0.14 & -0.07 & -0.02 & -0.04 & 0.05 & -0.14 & 0.38 \\ 
  & S.E. & 6.39 & 7.43 & 6.54 & 4.56 & 4.62 & 4.55 & 4.55 & 6.45 & 6.43 \\ 
\multirow{-3}{*}{OM1PM2}  & RMSE& 7.36 & 7.45 & 6.54 & 4.56 & 4.61 & 4.55 & 4.54 & 6.45 & 6.44 \\ 
\cline{2-11}
  & Bias& 1.18 & -0.71 & -0.34 & -0.18 & -0.16 & -0.13 & -0.05 & -0.37 & 0.17 \\ 
  & S.E. & 6.04 & 7.67 & 6.57 & 4.37 & 4.39 & 4.30 & 4.37 & 6.44 & 6.15 \\ 
\multirow{-3}{*}{OM2PM2}  & RMSE& 6.15 & 7.69 & 6.58 & 4.37 & 4.39 & 4.30 & 4.36 & 6.44 & 6.15 \\ 
   \hline
\end{tabular}
\end{table}

		Figure~\ref{fig:clear} shows the results of the simulation study above, where the dashed line represents the true parameter $\theta=E(Y)$. Table~\ref{tab:clear} presents the corresponding bias, standard error, and root mean square error for each method. Since the consistency of all estimators is guaranteed under OM1PM1, we can see that all estimators give similar performances except for the mean of the complete cases method. However, under OM1PM2, the consistency of the generalized linear model method is no longer guaranteed, while that of other estimators is still valid. In OM1PM2, the covariate balancing condition becomes critical, and so all methods satisfying the covariate balancing will be unbiased.  In OM2PM1, the proposed augmented PSW  estimator is still consistent, as the propensity score model is correctly specified. Our proposed augmented PSW  estimator with $\gamma$-divergence  has relatively lower root mean square error compared to other estimators. Our proposed augmented PSW  estimator with $\gamma$-divergence is biased in this scenario due to the misspecification of the outcome model, where other estimators are built under the correct specified response model. 
		In OM2PM2,  our proposed augmented PSW  estimator with $\gamma$-divergence is robust against other estimators, with fewer outliers and the smallest root mean square error.

	We also investigated the performance of the proposed linearization variance estimator in  \eqref{linearized variance estimation}. We computed confidence intervals based on asymptotic normality. The results are presented in Table~\ref{tab: Variance Estimation OM1PM1}. The performances are satisfactory when the ourcome regression model is specified correctly.

In addition, we check the robustness of the augmented propensity score weighting estimator with $\gamma$-divergence against outliers. After the data are generated under the same $2 \times 2$ factorial design as in the previous simulation setting, 
additional noise eight times a random variable generated  from  Student's t distribution with degree freedom of 8 is added to 20\% of the observed outcomes.   
Again, the sample size is $100$ and the Monte Carlo sample size is $1,000$.

\begin{table}[ht]
	\caption{Linearized variance estimation for gamma divergence method  where the errors are \textit{i.i.d.} from  standard Gaussian distribution}
	\label{tab: Variance Estimation OM1PM1}
	\centering
	\begin{tabular}{cccccc}
	\hline
		&  & \multicolumn{4}{c}{Method}\\
		\cline{3-6}
		\multirow{-2}{*}{Model}  & \multirow{-2}{*}{Nominal coverage rate} & $\rm{APS}_{0.3}$& $\rm{APS}_{0.5}$ & $\rm{APS}_{0.7}$ & $\rm{APS}_{1.0}$  \\ 
		\hline
&90\%  & 88.5 \% & 89.3 \% & 89.0 \% & 90.4 \% \\ 
\multirow{-2}{*}{OM1RM1} &95\%  & 94.2 \% & 94.5 \% & 94.4 \% & 96.0 \% \\ 
\cline{2-6}
&90\%  & 89.2 \% & 89.8 \% & 90.6 \% & 90.3 \% \\ 
\multirow{-2}{*}{OM1RM2} &95\%  & 94.5 \% & 94.8 \% & 95.2 \% & 95.1 \% \\ 
   \hline
\end{tabular}
\end{table}

Figure~\ref{fig:contaminated} shows the performance of various estimators, where the dashed line represents the true mean of $y_{i}'s$. Table~\ref{tab:contaminated} presents the corresponding bias, standard error, and root mean square error for each method.  Compared to other estimators, in all four scenarios, our proposed augmented propensity score weighting estimator with $\gamma$-divergence
apparently gives the smallest root mean square error, which validates our robustness claim in Section~\ref{section:RI_info_proj}.

\subsection{Real data application}

We further check our proposed estimators for artificial missingness with real data from
the California API program (http://api.cde.ca.gov/ or survey package \citep{surveypackage} in R \citep{R}). Within the data set, standardized student tests are performed to calculate the
API for California schools. 
Specifically, we select the API for year 2000 (api00) as the response variable. For covariates, we set the
API for year 1999 (api99) as $X_{1}$, the percentage of students eligible for subsidized meals (meals) as $X_{2}$, 
the percentage of English language learners (ell) as $X_{3}$, the average level of parental education (avg) as $X_{4}$,
the percentage of fully qualified teachers (full) as $X_{5}$, and the number of students enrolled (enroll) as $X_{6}$,
where the words in parentheses are the abbreviations for variable names in the dataset. 
We artificially created the missingness with the following two response mechanisms: 
PM3, where $\delta \mid  {\bX}$ follows the Bernoulli distribution with
$
\mathrm{logit}\{P(\delta = 1 \mid {\bX})\} =   \phi_0 + 2 X_{1} + X_{2} + 0.5 X_{3},
$
and $\phi_0$ is chosen to achieve the 60\% response rates;
PM4, where $\delta \mid {\bX}$ follows the Bernoulli distribution with
$
{P}( \delta=1 \mid {\bX} ) = 
0.8$, if  $a + 2 X_1 + X_2 + X_3 + X_4 + X_5+ X_6 > 0$,  
$
{P}( \delta=1 \mid {\bX} ) = 
0.4$ otherwise, 
and $a$ is chosen to achieve the response rates 60\%.

\begin{figure}
	\centering
	\begin{subfigure}{.3\linewidth}
		\caption{PM3}
		\includegraphics[scale=0.3]{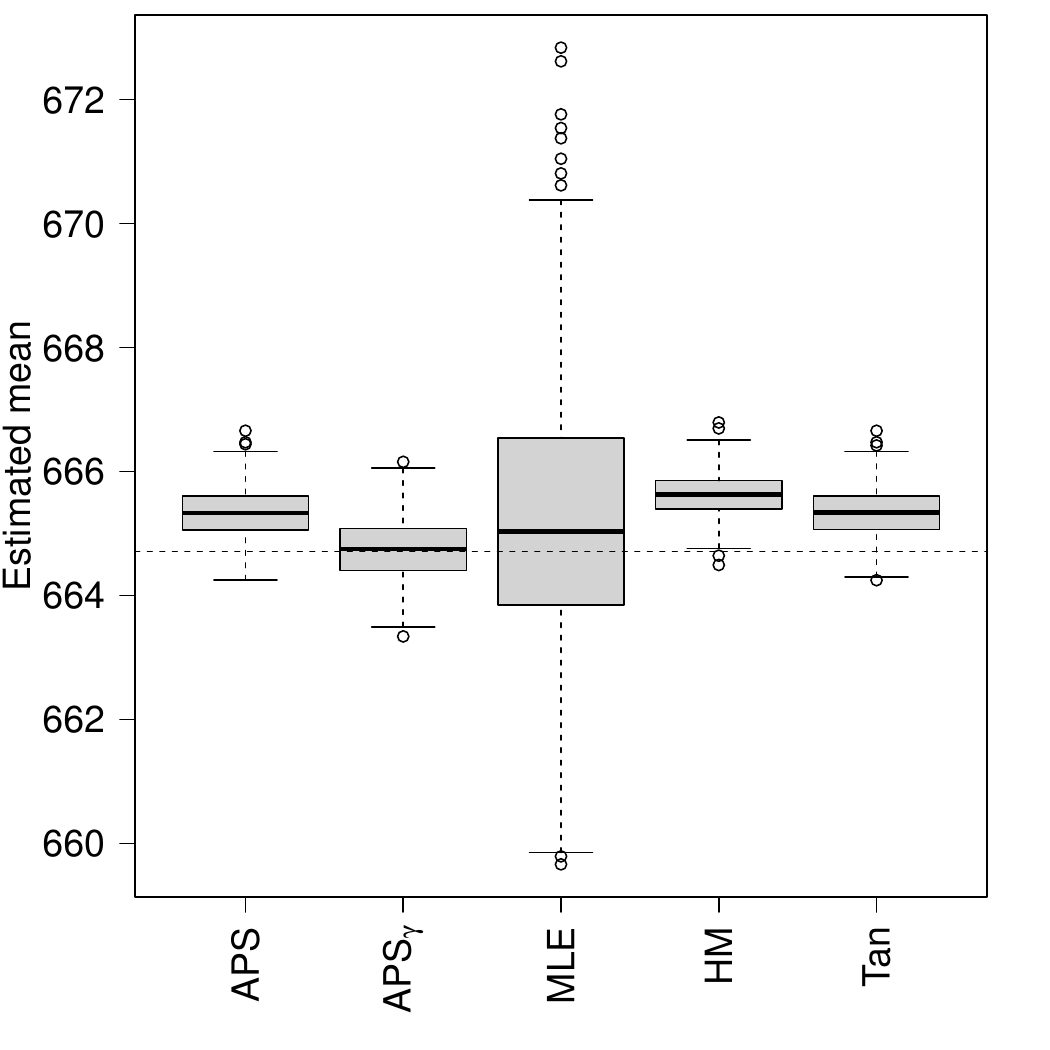}
	\end{subfigure}
	\hskip6em
	\begin{subfigure}{.3\linewidth}
		\caption{PM4}
		\includegraphics[scale=0.3]{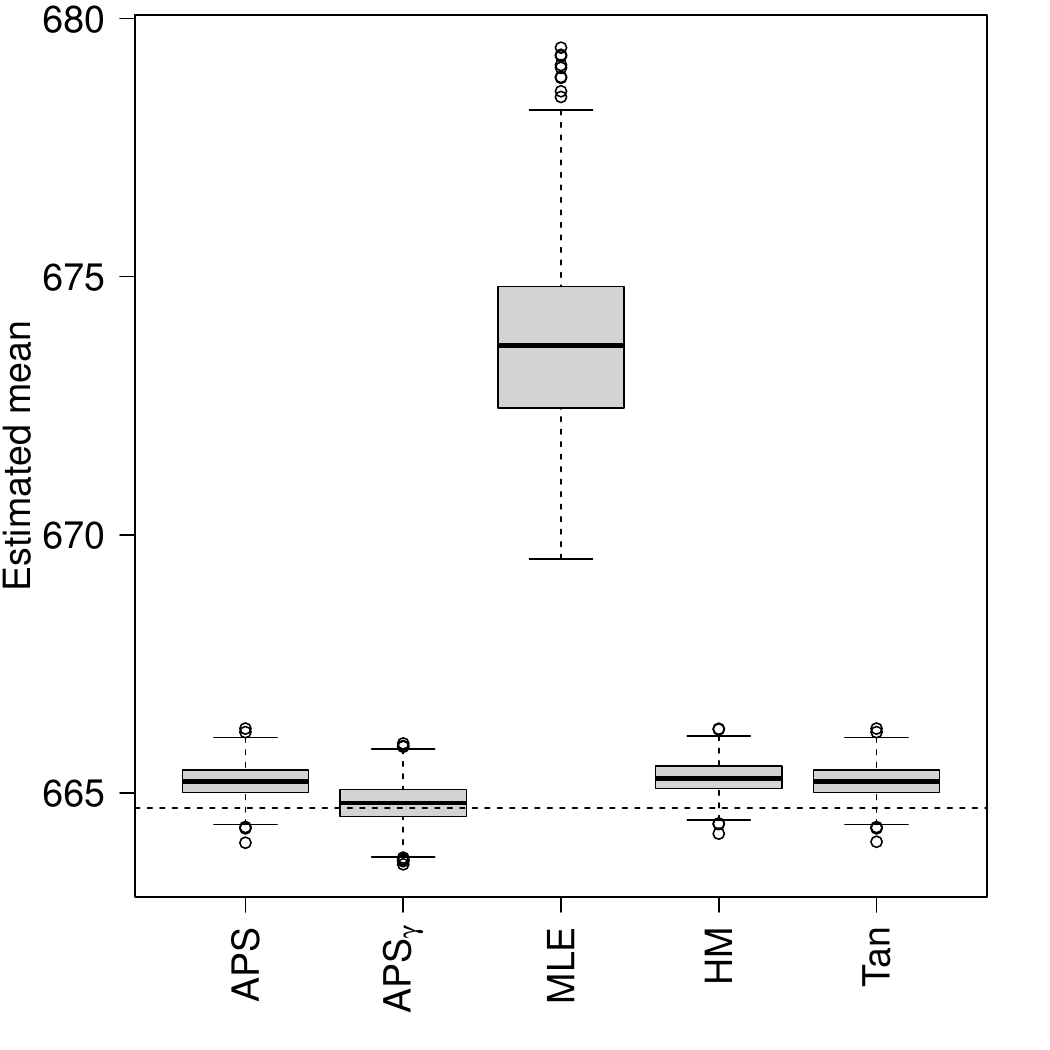}
	\end{subfigure}
	\caption{Estimators comparison in real data with $1000$ Monte Carlo sample, with artificially missingness mechanism PM3 and PM4.  APS: augmented propensity score weighting estimator; $\mathrm{APS}_{\gamma}$: augmented propensity score weighting with $\gamma$-divergence estimator, where $\gamma$ is chosen by the cross-validation method based on MSPE criterion;  MLE: maximum likelihood estimation; HM: entropy balancing method in Hainmueller (2012); Tan: regularized calibration method in Tan (2019).}
	\label{fig:real data}
\end{figure}

We compare the same estimators as in previous subsections with 1000 Monte Carlo samples. 
In particular, the mean of the complete cases method does not behave well in both cases and will affect the scale of the resultant box plots, so we
do not show its performance. Furthermore, we adopt the MSPE as the 5-fold cross-validation criterion for the selection strategy of $\gamma$, as described in Section~\ref{section:RI_info_proj}. 
For each Monte Carlo sample, we select the best $\gamma$ from $\{0.1, 0.2, \ldots, 1\}$ and denote the corresponding estimator as ${\mathrm{APS}}_{\gamma}$.
The results are summarized with box plots in Figure~\ref{fig:real data}, where the dashed line denotes the true population mean.  As we can see, the estimator ${\mathrm{APS}}_{\gamma}$ always gives the unbiased estimate with a relatively low root mean square error, outperforming other estimators.

\section{Application to CEAP data}
\label{section: CEAP_application}

	The Conservation Effects Assessment Project (CEAP) is a program initiated by the United States Department of Agriculture (USDA) Natural Resources Conservation Service (NRCS). CEAP collects and analyzes data from various sources, including field studies, monitoring sites, and modeling, to evaluate the impact of water and wind erosion.
Further details of the CEAP data can be found in \citet{berg2019semiparametric}.
The farmer's interview data together with the NRI data serve as input to the revised Universal Soil Loss Equation (RUSLE2) to generate a measure of sheet and rill erosion. In addition, RUSLE2 is also an advancement of a traditional approximation called the
Universal Soil Loss Equation (USLE). For the CEAP sample, the RUSLE2 suffers from missingness due to the refusal of the farmer interviewed, while the corresponding USLE is available.

\begin{figure}
	\centering
	\begin{subfigure}{.35\linewidth}
		\caption{Normal Q-Q plot}
		\includegraphics[width = 4cm, height = 4cm]{state_5.eps}
	\end{subfigure}
	\hskip2em
	\begin{subfigure}{.35\linewidth}
		\caption{Mean estimation}
		\includegraphics[width = 5.28cm, height = 4cm]{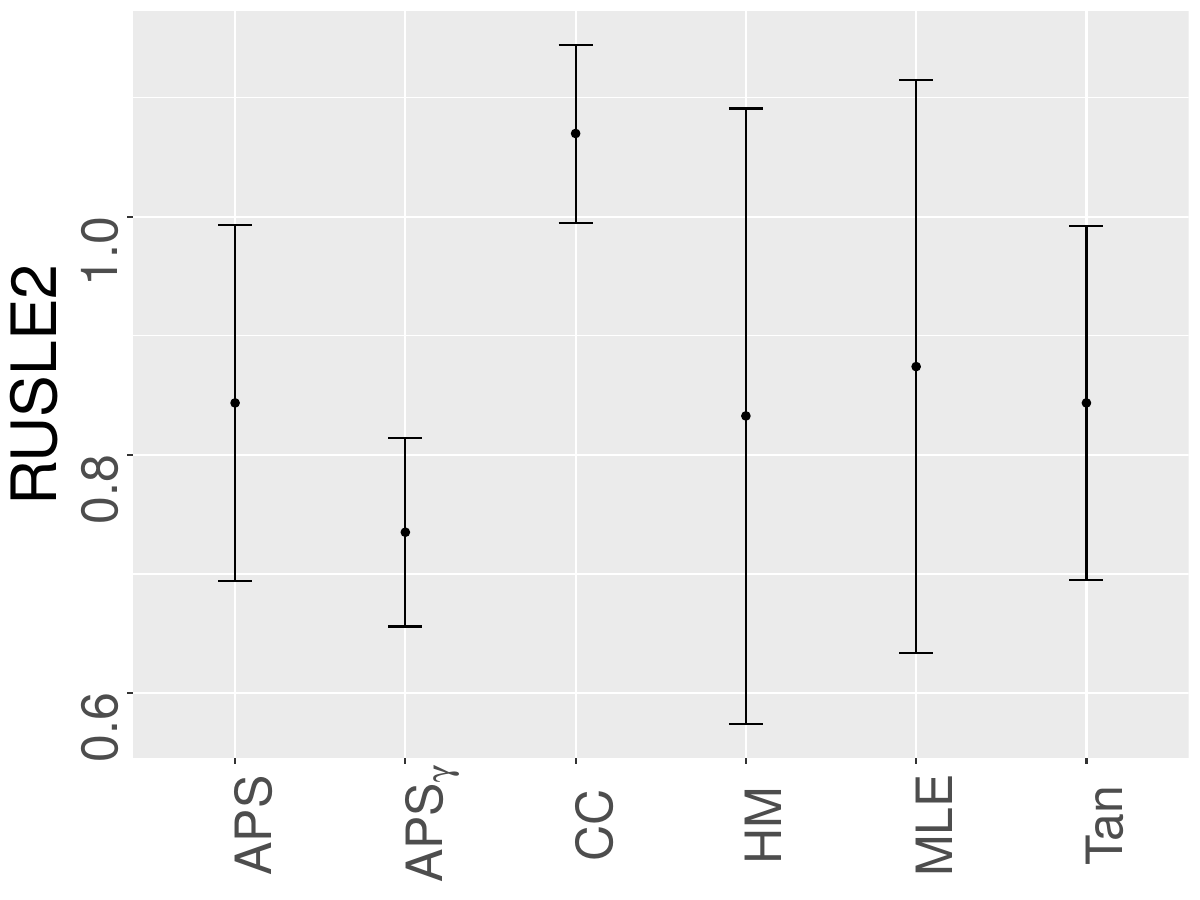}
	\end{subfigure}
	\caption{Quantile-quantile plot and mean estimation with 95\% confidence band for CEAP data in Arkansas.}
	\label{fig:real data CEAP}
\end{figure}

We are interested in estimating the population mean of RUSLE2 using USLE as an auxiliary variable for calibration weighting in Arkansas. Specifically, the total sampled points are 1509 and the fully observed are 406, with observed rate $26.9\%$. 
The normal quantile-quantile (Q-Q) plot generated from linear models based on the complete cases is presented in Sub-figure (a) of  Figure~\ref{fig:real data CEAP}. By the normal Q-Q plot,  
numerous outlier data points are evident, and the residuals deviate from satisfying the assumption of a normal distribution. Therefore, given this scenario, we have strong reasons to place greater trust in our suggested robust approach. The associated mean estimation result is presented in Sub-figure (b) of  Figure~\ref{fig:real data CEAP}, where the confidence 95\% confidence intervals for the APS method is constructed by \eqref{33} and \eqref{influence} and the $\rm{APS}_{\gamma}$ is constructed by \eqref{linearized variance estimation} and \eqref{estimated influence robust}. In this presentation, our proposed method using gamma divergence exhibits a narrow 95\% confidence interval and yields a low RUSLE2 value.

\section{Concluding Remarks}
\label{section: conclusion}

We have applied the information projection  to obtain an augmented PS model that allows for doubly robust estimation.  We have introduced self-efficiency to obtain the algebraic equivalence between the PSW estimator and the regression imputation estimator.  In addition, $\gamma$-power divergence can be used to obtain an outlier-robust regression imputation estimator, which can be expressed as a PSW estimator under self-efficiency.  In practice, an efficient and outlier-robust PSW estimator is very attractive  as the existence of outliers can often damage the efficiency of the result estimator. 
The tuning parameter is determined to balance the trade-off between statistical efficiency and robustness against outliers. In the future, further theoretical investigation on the effect of the choice of parameter $\gamma$ on the final estimation will be considered.

There are several directions for further extensions of the proposed method.  The proposed method is directly applicable to the calibration weighting in survey sampling \citep{fuller09, tille2020}.   
Other divergence measures such as Hellinger divergence  \citep{etho2021} 
can be also considered in the information projection. 
The proposed method is based on the assumption of missing at random. Extension to nonignorable nonresponse can also be an interesting research direction. Furthermore, the proposed method can be used for causal inference, including the estimation of the average treatment effect from observational studies \citep{yangding2020, chen2021multiple, chen2023effect}.  Developing tools for causal inference using the proposed  method will be an important extension of this research.

\appendix

\section{Proof of Lemma~\ref{lemma1}}
If the balancing condition \eqref{balancing} holds, then the propensity score weighting  estimator can be expressed as
\begin{eqnarray*}
\hat{\theta}_{\rm{PSW}} &=& \frac{1}{n} \sum_{i=1}^{n} \delta_{i} \omega_{i} y_{i} \\
&=& \frac{1}{n} \sum_{i=1}^{n} \delta_{i} \omega_{i} \left ( m_{i} + \varepsilon_{i} \right ) \\
&=& \frac{1}{n} \sum_{i=1}^{n} \delta_{i} \omega_{i} m_{i} + \frac{1}{n} \sum_{i=1}^{n} \delta_{i} \omega_{i} \varepsilon_{i} \\
&=& \frac{1}{n} \sum_{i=1}^{n} m_{i} + \frac{1}{n} \sum_{i=1}^{n} \delta_{i} \omega_{i} \varepsilon_{i},
\end{eqnarray*}
where $m_{i} = m(\B{x}_{i}; \B{\beta})$. Then 
	\begin{equation*}
	\hat{\theta}_{\rm PSW} - \hat{\theta}_{\rm{com}}
	=  \frac{1}{n}\sum_{i=1}^n \left( \delta_i \omega_i -1  \right) \varepsilon_i . \end{equation*}
	Under the MAR assumption, we can obtain 
	$$ \E \left (  \hat{\theta}_{\rm PSW} - \hat{\theta}_{\rm{com}}  \mid \B{x}_{1}, \ldots, \B{x}_{n}, \B{\delta} \right ) 
	= \frac{1}{n}\sum_{i=1}^n \left( \delta_i \omega_i -1  \right) \E( \varepsilon_i \mid \bx_i )  =0. $$
	Thus, we can conclude that $\hat{\theta}_{\rm PSW}$ is unbiased for $\theta$.

	\section{Proof of Lemma~\ref{lem2}}

Note that we can express the covariate-balancing condition in (\ref{balancing})  as 
\begin{equation} 
\sum_{i=1}^n \left ( 1- \delta_i \omega_i \right ) \bb_{i} = \mathbf{0}.
\label{dw}
\end{equation} 
Thus, as long as (\ref{dw}) is satisfied, we can write  
\begin{eqnarray*}
\hat{\theta}_{\rm PSW} &=& n^{-1} \sum_{i =1}^n \delta_i \omega_i y_i + n^{-1} \sum_{i=1}^n \left ( 1- \delta_i \omega_i \right ) \B{\OP{b}}_{i}^{\T} \B{\beta}\\
&=& n^{-1} \sum_{i=1}^n \left\{ \delta_i y_i + (1- \delta_i) \B{\OP{b}}_{i}^{\T} \B{\beta}  \right\}  + n^{-1}  \sum_{i=1}^n \delta_i (  \omega_i -1  )   ( y_i - \B{\OP{b}}_{i}^{\T} \B{\beta}  )
\end{eqnarray*} 
for any $\bbeta$. Thus, as long as (\ref{ibc-2}) hold, we can  obtain (\ref{eq7}).

\section{Proof of Theorem~\ref{T1}}

Since $\hat{\blambda}$ satisfies (\ref{est3}), we can obtain the following equivalence   
\begin{eqnarray*} 
\hat{\theta}_{\rm APSW}  &=& \frac{1}{n} \sum_{i=1}^n \delta_i \omega^*  ( \B{x}_i; \hat{\bphi},  \hat{\blambda} ) y_i - \bU_{2, n}^{\T} ( \hat{\blambda} \mid \hat{\bphi}  ) \bbeta \\
&=& \frac{1}{n} \sum_{i=1}^n \left\{ \bb_i^{\T} \bbeta + \delta_i \omega^* ( \B{x}_i; \hat{\bphi},  \hat{\blambda} ) ( y_i - \bb_i^{\T} \bbeta ) \right\} \\
&:=& \hat{\theta}_{\rm APSW} ( \hat{\blambda}; \bbeta )  
\end{eqnarray*}
for any $\bbeta$. Now, 
\begin{eqnarray*}
\frac{ \partial}{ \partial \blambda } \hat{\theta}_{\rm APSW}  ( \blambda; \bbeta ) &=& \frac{1}{n} \sum_{i=1}^n \delta_i  \left( \hat{\pi}_{1,i}^{-1}-1\right) \exp \left( \bb_i^{\T} \blambda \right) \left ( y_i - \bb_i^{\T} \bbeta \right ) \bb_i .
\end{eqnarray*} 
Thus,  for the choice of $\hat{\bbeta}$ in 
(18), we can obtain 
$$ \E\left\{ \frac{ \partial}{ \partial \blambda } \hat{\theta}_{\rm APSW} ( \blambda; {\bbeta}^* ) \right\}=\B{0}. $$
Thus, the uncertainty associated with $\hat{\blambda}$ is asymptotically negligible at $\bbeta= \bbeta^*$. Thus, we obtain 
\begin{equation}
\hat{\theta}_{\rm{APSW}} ( \blambda^{\ast}; \bbeta^{\ast}  ) =\frac{1}{n} \sum_{i=1}^n \left\{ \B{\OP{b}}_i^{\T} {\bbeta}^* + \delta_i \hat{\omega}^*  ( \bx_i; \hat{\bphi}, \blambda^*  ) ( y_i - \bb_i^{\T} {\bbeta}^*  ) \right\} +o_p(n^{-1/2}). \label{ee2}
\end{equation}
Now, to apply Taylor linearization with respect to $\bphi$, we define 
\begin{eqnarray*}
\hat{\theta}_{\ell}  ( \hat{\bphi}; \B{\kappa}  )  &\equiv &  \frac{1}{n} \sum_{i=1}^n \left\{ \bb_i^{\T} {\bbeta}^* + \delta_i \hat{\omega}^*  ( \bx_i; \hat{\bphi}, \blambda^* ) ( y_i - \bb_i^{\T} {\bbeta}^*  ) \right\} - \bU_{1, n}^{\T}  ( \hat{\bphi}  ) \B{\kappa} \\
&=& \frac{1}{n} \sum_{i=1}^n \left[ \delta_i y_i + (1-\delta_i) \left ( \bb_i^{\T} {\bbeta}^* + \bh_i^{\T} \B{\kappa} \right ) \right.\notag\\
& &\qquad\qquad\left.+ \delta_i \left( \hat{\pi}_{1, i}^{-1} -1 \right)\left\{ \exp (\bb_i^{\T} \blambda^*) ( y_i - \bb_i^{\T} {\bbeta}^*)- \bh_i^{\T} \B{\kappa}  \right\}\right].
\end{eqnarray*}

Thus, we can choose $\B{\kappa}^*$ to satisfy 
$$ \E\left\{ \frac{\partial}{\partial \bphi} \hat{\theta}_{\ell} ({\bphi}; \B{\kappa}^*) \Big |_{\bphi = \bphi^{\ast}} \right\}= \mathbf{0}, $$
which gives the final linearization form as in (\ref{33}).

\section{Regularity Conditions for Theorem~\ref{TR}}\label{app: regularity robust}

Before the statement of regularity conditions for Theorem~\ref{TR}, we first define a few quantities. 
Let
\begin{align}
&\bU_{3, n}(\blambda, \bbeta, \sigma^{2}\mid \bphi^{\ast}) =\notag\\
&\quad\begin{pmatrix}
\frac{1}{n}\left[\sum_{i=1}^{n} \delta_{i}\left\{
		1 +   \{d_{i}(\bphi^{\ast}) - 1\}g_{i}(\blambda)
		q_{\gamma, i}(\bbeta, \sigma^{2})
	\right\}\bb_{i}
-\sum_{i=1}^{n}\bb_{i}\right]\\
\frac{1}{n}\sum_{i=1}^{n}\delta_{i} \{d_{i}(\bphi^{\ast}) - 1 \}g_{i}(\blambda) 
q_{\gamma, i}(\bbeta, \sigma^{2})(y_{i} - \bb_{i}\trans\bbeta)\bb_{i}\\
\frac{1}{n} \sum_{i=1}^{n}\delta_{i}\{d_{i}(\bphi^{\ast}) - 1 \}g_{i}(\blambda) 
q_{\gamma, i}(\bbeta, \sigma^{2})(y_{i} - \bb_{i}\trans\bbeta)^{2} 
\left\{
	(y_{i} - \bb_{i}\trans\bbeta)^{2} - \frac{\sigma^{2}}{1  + \gamma}
\right\}
\end{pmatrix}.
\end{align}
Further, define

$$
\bs_{11} = -\sum_{i=1}^{n} \delta_{i}\left\{
	\{d_{i}(\wh{\bphi}) - 1\}g_{i}(\blambda)
	q_{\gamma, i}(\bbeta, \sigma^{2})
	\right\}\bb_{i}\bb_{i}\trans,
$$
$$
\bs_{12}= \sum_{i=1}^{n}\delta_{i}
\{d_{i}(\wh{\bphi}) - 1\}g_{i}(\blambda)
q_{\gamma, i}(\bbeta, \sigma^{2})(y_{i} - \bb_{i}\trans\bbeta)\bb_{i}\bb_{i}\trans,
$$

$$
\bs_{13}= \sum_{i=1}^{n}\delta_{i}\{d_{i}(\wh{\bphi}) - 1\}g_{i}(\blambda)
q_{\gamma, i}(\bbeta, \sigma^{2})(y_{i} - \bb_{i}\trans\bbeta)^{2} 
\left\{
	(y_{i} - \bb_{i}\trans\bbeta)^{2} - \frac{ \sigma^{2}}{1 + \gamma}
\right\}\bb_{i},
$$

$$
\bs_{21} = -\sum_{i=1}^{n}\delta_{i}\{d_{i}(\wh{\bphi}) - 1\}g_{i}(\blambda)
q_{\gamma, i}(\bbeta, \sigma^{2})
\left\{
\frac{\gamma}{\sigma^{2}}(y_{i} - \bb_{i}\trans\bbeta)
\right\}
\bb_{i}\bb_{i}\trans,
$$
$$\bs_{22} = \sum_{i=1}^{n}\delta_{i}\{d_{i}(\wh{\bphi}) - 1\}g_{i}(\blambda)
q_{\gamma, i}(\bbeta, \sigma^{2})
\left\{
\frac{\gamma}{\sigma^{2}}(y_{i} - \bb_{i}\trans\bbeta)^{2} -1 
\right\}
\bb_{i}\bb_{i}\trans,
$$
$$
\bs_{23} = \sum_{i=1}^{n}\delta_{i}\{d_{i}(\wh{\bphi}) - 1\}g_{i}(\blambda)
q_{\gamma, i}(\bbeta, \sigma^{2})(y_{i} - \bb_{i}\trans\bbeta)
\left\{
\frac{\gamma}{\sigma^{2}}(y_{i} - \bb_{i}\trans\bbeta)^{2} - 2 
\right\}
\bb_{i},
$$

$$
\bs_{31} = - \sum_{i=1}^{n}\delta_{i}\{d_{i}(\wh{\bphi}) - 1\}g_{i}(\blambda)
q_{\gamma, i}(\bbeta, \sigma^{2})\frac{\gamma}{2(\sigma^{2})^{2}}(y_{i} - \bb_{i}\trans\bbeta)^{2}\bb_{i}\trans,
$$
$$
\bs_{32} = \sum_{i=1}^{n}\delta_{i}\{d_{i}(\wh{\bphi}) - 1\}g_{i}(\blambda)
q_{\gamma, i}(\bbeta, \sigma^{2})\frac{\gamma}{2(\sigma^{2})^{2}}(y_{i} - \bb_{i}\trans\bbeta)^{3}\bb_{i}\trans,
$$
\begin{align*}
s_{33} =& \sum_{i=1}^{n}\delta_{i}\{d_{i}(\wh{\bphi}) - 1\}g_{i}(\blambda)
q_{\gamma, i}(\bbeta, \sigma^{2}) \notag\\
&\qquad\qquad\times \left[
	\frac{\gamma}{2 (\sigma^{2})^{2} } (y_{i} -\bb_{i}\trans\bbeta)^{2}
	\left\{
	(y_{i} -\bb_{i}\trans\bbeta)^{2} - \frac{\sigma^{2}}{1 + \gamma}
	\right\}
	- \frac{1}{  1 + \gamma  }
\right].
\end{align*}
Let  
$$
\bS_{n}(\blambda, \bbeta, \sigma^{2}\mid \wh{\bphi}) = \begin{pmatrix}
\bs_{11} & \bs_{12} & \bs_{13}  \\
\bs_{21} & \bs_{22} & \bs_{23}  \\
\bs_{31} & \bs_{32} & \bs_{33}
\end{pmatrix}.
$$

In addition with the regularity conditions in Theorem 1, we need the following additional assumptions.

\begin{assumption}\label{unique solution robust 0}
  The estimating equation   $\bU_{3, n}(\blambda, \bbeta, \sigma^{2}  \mid \bphi^{\ast}) = \bzero$  has a unique solution $(\wh{\blambda}, \wh{\bbeta}_{q}, \wh{\sigma}_{q}^{2})$ and there exists a function $\bU_{3}(\blambda, \bbeta, \sigma^{2}  \mid \bphi^{\ast})$ such that $\bU_{3, n}(\blambda, \bbeta, \sigma^{2}  \mid \bphi^{\ast}) \rightarrow \bU_{3}(\blambda, \bbeta, \sigma^{2}  \mid \bphi^{\ast})$ uniformly as $n \rightarrow \infty$ and $\bU_{3}(\blambda, \bbeta, \sigma^{2} \mid \bphi^{\ast}) = \bzero$ has a unique solution $(\blambda^{\ast}, \bbeta^{\ast}, \sigma^{\ast 2})$.  
    \end{assumption}

\begin{assumption}\label{unique solution robust 1}
  The limiting function $\bU_{3}(\blambda, \bbeta, \sigma^{2}  \mid \bphi^{\ast})$ is differentiable and $\partial_{ (\blambda\trans, \bbeta\trans, \sigma^{2} )\trans  }\bU_{3}(\blambda, \bbeta, \sigma^{2}  \mid \bphi^{\ast})$ is continuous on a compact set $\mathcal{G}_{3}$ containing $(\blambda^{\ast}, \bbeta^{\ast}, \sigma^{\ast 2})$. 
  \end{assumption}

\begin{assumption}\label{unique solution robust 2} The matrix $ \partial_{ (\blambda\trans, \bbeta\trans, \sigma^{2} )\trans  } \bU_{3}(\blambda, \bbeta, \sigma^{2}\mid \bphi^{\ast}) 
  \mid_{\blambda = \blambda^{\ast}, \bbeta = \bbeta^{\ast}, \sigma^{2} = \sigma^{\ast 2} }$
is non-singular.
\end{assumption}

\begin{assumption}\label{existence robust}
  There exists a function $\bS(\bphi)$ such that $\bS_{n}(\blambda, \bbeta, \sigma^{2}\mid \bphi^{\ast}) \rightarrow \bS(\blambda, \bbeta, \sigma^{2}\mid \bphi^{\ast})$ uniformly as $n \rightarrow \infty$ and $\bS(\blambda^*, \bbeta^*, \sigma^{*2}\mid \bphi^{\ast})$ is non-singular.
  \end{assumption}

Assumptions~\ref{unique solution robust 0} -- \ref{unique solution robust 2} are commonly used in estimating  equation theory and also ensure the existence of $\blambda^*$, $\bbeta^*$ and $\sigma^{*2}$. Assumption~\ref{existence robust} ensures the existence of the nuisance parameter $\mu^*, \bzeta^*$ and $\nu^*$ defined in  Theorem~\ref{TR}, which helps us represent the influence function for the proposed robust estimator.

 \section{Proof of Theorem~\ref{TR}}\label{appendix:E}

 By the normal equations, we have 
 \begin{align}
 &\theta_{\rm APSW, \gamma}( \wh{\blambda}, \wh{\bbeta}_{q},  \wh{\sigma}_{q}^{2}\mid \wh{\bphi};\bmu, \bzeta, \nu)\notag\\
 =& 
 \frac{1}{n}\sum_{i=1}^{n}\delta_{i}
 \omega_{\gamma,i}(\bx_{i}; \wh{\blambda}, \wh{\bbeta}_{q}, \wh{\sigma}_{q}^{2}, \wh{\bphi})y_{i}   \notag\\
 &\quad - \frac{\bmu\trans}{n}
 \left[
   \sum_{i=1}^{n} \delta_{i}\left\{
     1 +   \{d_{i}(\wh{\bphi}) - 1\}g_{i}(\wh{\blambda})
     q_{\gamma, i}(\wh{\bbeta}_{q}, \wh{\sigma}_{q}^{2})
   \right\}\bb_{i}
 -\sum_{i=1}^{n}\bb_{i}
 \right]\notag\\
 &\quad+ \frac{\bzeta\trans}{n}\sum_{i=1}^{n}\delta_{i}
 \{d_{i}(\wh{\bphi}) - 1\}g_{i}(\wh{\blambda})
 q_{\gamma, i}(\wh{\bbeta}_{q}, \wh{\sigma}_{q}^{2})(y_{i} - \bb_{i}\trans\wh{\bbeta}_{q})\bb_{i}\notag\\
 &\quad + \frac{\nu}{n} \sum_{i=1}^{n}\delta_{i}\{d_{i}(\wh{\bphi}) - 1\}g_{i}(\wh{\blambda})
 q_{\gamma, i}(\wh{\bbeta}_{q}, \wh{\sigma}_{q}^{2})(y_{i} - \bb_{i}\trans\wh{\bbeta}_{q})^{2} 
 \left\{
   (y_{i} - \bb_{i}\trans\wh{\bbeta}_{q})^{2} - \frac{ \wh{\sigma}_{q}^{2}}{1 + \gamma}
 \right\}\notag\\
 =& \frac{1}{n}\sum_{i=1}^{n}\bb_{i}\trans \bmu + \frac{1}{n}\sum_{i=1}^{n}
 \delta_{i} \omega_{\gamma,i}(\bx_{i}; \wh{\blambda}, \wh{\bbeta}_{q}, \wh{\sigma}_{q}^{2}, \wh{\bphi})(y_{i} - 
 \bb_{i}\trans \bmu)
 \notag\\
 &\quad+ \frac{\bzeta\trans}{n}\sum_{i=1}^{n}\delta_{i}
 \{d_{i}(\wh{\bphi}) - 1\}g_{i}(\wh{\blambda})
 q_{\gamma, i}(\wh{\bbeta}_{q}, \wh{\sigma}_{q}^{2})(y_{i} - \bb_{i}\trans\wh{\bbeta}_{q})\bb_{i}\notag\\
 &\quad + \frac{\nu}{n} \sum_{i=1}^{n}\delta_{i}\{d_{i}(\wh{\bphi}) - 1\}g_{i}(\wh{\blambda})
 q_{\gamma, i}(\wh{\bbeta}_{q}, \wh{\sigma}_{q}^{2})(y_{i} - \bb_{i}\trans\wh{\bbeta}_{q})^{2} 
 \left\{
   (y_{i} - \bb_{i}\trans\wh{\bbeta}_{q})^{2} - \frac{ \wh{\sigma}_{q}^{2}}{1 + \gamma}
 \right\}
   \end{align}
 for any $\bmu \in \mathbb{R}^{p}$, $\bzeta \in \mathbb{R}^{p}$ and  $\nu \in \mathbb{R}$. 
 In addition, tedious derivation will lead 
 \begin{align}
   \begin{pmatrix}
   \frac{\partial}{\partial \blambda} \theta_{\rm APSW, \gamma}(\bbeta, \sigma^{2}, \blambda\mid \wh{\bphi}; \bmu, \bzeta, \nu)\\
   \frac{\partial}{\partial \bbeta} \theta_{\rm APSW, \gamma}(\bbeta , \sigma^{2}, \blambda\mid \wh{\bphi}; \bmu, \bzeta, \nu)\\
   \frac{\partial}{\partial \sigma^{2}} \theta_{\rm APSW, \gamma}(\bbeta, \sigma^{2}, \blambda\mid \wh{\bphi};\bmu, \bzeta, \nu)
   \end{pmatrix}
   = \frac{1}{n}\begin{pmatrix}
   \bs_{10} + \bs_{11}\bmu + \bs_{12} \bzeta
 + \bs_{13} \nu  \\
   s_{20} + \bs_{21}\bmu +  \bs_{22} \bzeta + s_{23}\nu  \\
   \bs_{30} + \bs_{31}\bmu + \bs_{32} \bzeta + \bs_{33}\nu 
   \end{pmatrix}
  \label{F.2}
 \end{align}
 where 
 $$
 \bs_{10} = \sum_{i=1}^{n} \delta_{i}\left\{
   \{d_{i}(\wh{\bphi}) - 1\}g_{i}(\blambda)
   q_{\gamma, i}(\bbeta, \sigma^{2})y_{i}
   \right\}\bb_{i},
 $$
 
 $$
 \bs_{20} = \sum_{i=1}^{n}\delta_{i}\{d_{i}(\wh{\bphi}) - 1\}g_{i}(\blambda)
 q_{\gamma, i}(\bbeta, \sigma^{2})
 \left\{
 \frac{\gamma}{\sigma^{2}}(y_{i} - \bb_{i}\trans\bbeta)
 \right\}
 y_{i}\bb_{i},
 $$

 $$
 \bs_{30} =  \sum_{i=1}^{n}\delta_{i}\{d_{i}(\wh{\bphi}) - 1\}g_{i}(\blambda)
 q_{\gamma, i}(\bbeta, \sigma^{2})\frac{\gamma}{2(\sigma^{2})^{2}}(y_{i} - \bb_{i}\trans\bbeta)^{2}y_{i},
 $$
 and other quantities are defined in Appendix~\ref{app: regularity robust}.

Therefore, we can choose $\bmu^{\ast}$, $\bzeta^{\ast}$ and $\nu^{\ast}$  such that 
\begin{align}
	E \left\{
		\begin{pmatrix}
		\frac{\partial}{\partial \blambda} \theta_{\rm APSW, \gamma}(\blambda, \bbeta, \sigma^{2} \mid \wh{\bphi}; \bmu^{\ast}, \bzeta^{\ast}, \nu^{\ast})\\
			\frac{\partial}{\partial \bbeta} \theta_{\rm APSW, \gamma}(\blambda, \bbeta, \sigma^{2}\mid \wh{\bphi};  \bmu^{\ast}, \bzeta^{\ast}, \nu^{\ast})\\
			\frac{\partial}{\partial \sigma^{2}} \theta_{\rm APSW, \gamma}(\blambda, \bbeta, \sigma^{2}\mid \wh{\bphi};  \bmu^{\ast}, \bzeta^{\ast}, \nu^{\ast})
			\end{pmatrix}\Bigg{|}_{\substack{\blambda = \blambda^{\ast}, \\ \bbeta = \bbeta^{\ast},\\ \sigma^{2} = \sigma^{\ast 2} }}
	\right\} = \bzero,
\end{align}
where ensures that 
\begin{align}
		&\quad \theta_{\rm APSW, \gamma}(\blambda^{\ast}, {\bbeta}^{\ast}, \sigma^{\ast 2} \mid \wh{\bphi}; \bmu^{\ast}, \bzeta^{\ast}, \nu^{\ast}) \notag\\
	&=  \frac{1}{n}\sum_{i=1}^{n} \bb_{i}\trans \bmu^{\ast} + \frac{1}{n}\sum_{i=1}^{n}
\delta_{i} \omega_{\gamma,i}(\bx_{i}; \blambda^{\ast}, \bbeta^{\ast}, \sigma^{\ast 2}, \wh{\bphi})(y_{i} - 
\bb_{i}\trans \bmu^{\ast}) \notag\\
&\quad+ \frac{\bzeta^{\ast \rm{T}}}{n}\sum_{i=1}^{n}\delta_{i}
\{d_{i}(\wh{\bphi}) - 1\}g_{i}(\blambda^{\ast})
q_{\gamma, i}(\bbeta^{\ast}, \sigma^{\ast 2})(y_{i} - \bb_{i}\trans\bbeta^{\ast})\bb_{i}\notag\\
		&\quad + \frac{\nu^{\ast}}{n} \sum_{i=1}^{n}\delta_{i}\{d_{i}(\wh{\bphi}) - 1\}g_{i}(\blambda^{\ast})
		q_{\gamma, i}(\bbeta^{\ast}, \sigma^{\ast 2})(y_{i} - \bb_{i}\trans \bbeta^{\ast})^{2} 
		\left\{
			(y_{i} - \bb_{i}\trans \bbeta^{\ast})^{2} - \frac{\sigma^{\ast 2}}{ 1  + \gamma}
		\right\}\notag\\
		&=  \theta_{\rm APSW, \gamma}(\wh{\blambda}, \wh{\bbeta}_{q}, \wh{\sigma}_{q}^{2}, \wh{\bphi} ) + o_{p}(n^{-1/2}).
\end{align}
Finally, we have 
\begin{align}
	&\quad  \theta_{\rm APSW, \gamma}(\wh{\bphi}\mid \blambda^{\ast}, {\bbeta}^{\ast}, \sigma^{\ast 2};\bmu^{\ast}, \bzeta^{\ast}, \nu^{\ast}, \bkappa ) \notag\\ 
	&=   \frac{1}{n}\sum_{i=1}^{n} \bb_{i}\trans \bmu^{\ast} + \frac{1}{n}\sum_{i=1}^{n}
\delta_{i} \omega_{\gamma,i}(\bx_{i}; \blambda^{\ast}, \bbeta^{\ast}, \sigma^{\ast 2}, \wh{\bphi})(y_{i} - 
\bb_{i}\trans \bmu^{\ast}) \notag\\
&\quad+
	\frac{\bzeta^{\ast {\rm T}}}{n} \sum_{i=1}^{n}\delta_{i}\{d_{i}(\wh{\bphi}) - 1\}g_{i}(\blambda^{\ast})
q_{\gamma, i}(\bbeta^{\ast}, \sigma^{\ast 2})(y_{i} - \bb_{i}\trans\bbeta^{\ast})
\bb_{i}\notag\\
&\quad + \frac{\nu^{\ast}}{n} \sum_{i=1}^{n}\delta_{i}\{d_{i}(\wh{\bphi}) - 1\}g_{i}(\blambda^{\ast})
q_{\gamma, i}(\bbeta^{\ast}, \sigma^{\ast 2})(y_{i} - \bb_{i}\trans \bbeta^{\ast})^{2} 
\left\{
	(y_{i} - \bb_{i}\trans \bbeta^{\ast})^{2} - \frac{\sigma^{\ast 2}}{ 1 + \gamma}
\right\} \notag\\
&\quad 
- \bU_{1, n}\trans(\wh{\bphi})\bkappa
\end{align}
holds for any $\bzeta \in \mathbb{R}^{p}$.
Apparently, 
\begin{align}
	&\quad\frac{\partial}{\partial \bphi}\theta_{\rm APSW, \gamma}(\bphi\mid\blambda^{\ast},  {\bbeta}^{\ast}, \sigma^{\ast 2}; \bmu^{\ast}, \bzeta^{\ast}, \nu^{\ast}, \bkappa)
	\notag\\
	&=  - \frac{1}{n}\sum_{i=1}^{n}
		\delta_{i} d_{i}^{2}(\bphi) 
	g_{i}(\blambda)q_{\gamma, i}(\bbeta^{\ast}, \sigma^{\ast 2})
	(y_{i} - \bb_{i}\trans\bmu^{\ast}  )\frac{\partial}{\partial \bphi}\pi_{1}^{(0)}(\bx_{i}, \bphi)\notag\\
	&\quad - \frac{1}{n}\sum_{i=1}^{n}
		\delta_{i} d_{i}^{2}(\bphi) 
	g_{i}(\blambda)q_{\gamma, i}(\bbeta^{\ast}, \sigma^{\ast 2})(y_{i} - \bb_{i}\trans\bbeta^{\ast})
	(\bzeta^{\ast {\rm T}}\bb_{i})\frac{\partial}{\partial \bphi}\pi_{1}^{(0)}(\bx_{i}, \bphi)\notag\\
	&\quad - \frac{\nu^{\ast}}{n} \sum_{i=1}^{n} 
		\delta_{i}d_{i}^{2}(\bphi) 
	g_{i}(\blambda)q_{\gamma, i}(\bbeta^{\ast}, \sigma^{\ast 2})(y_{i} - \bb_{i}\trans \bbeta^{\ast})^{2} 
	\notag\\
 &\qquad\qquad\qquad\times\left\{
		(y_{i} - \bb_{i}\trans \bbeta^{\ast})^{2} - \frac{\sigma^{\ast 2}}{ 1 + \gamma}
	\right\} \frac{\partial}{\partial \bphi}\pi_{1}^{(0)}(\bx_{i}, \bphi)\notag\\
		&\quad-
	\frac{1}{n}\sum_{i=1}^{n}\left[
		\left\{
		\delta_{i}d_{i}(\bphi)  -1
		\right\}
  \left\{\frac{\partial\bh(\bx_{i};\bphi)}{\partial \bphi}\right\}\trans
	- \delta_{i}d_{i}^{2}(\bphi) 
	\frac{\partial \pi_{1}^{(0)}(\bx_{i}, \bphi)}{\partial \bphi} \bh\trans(\bx_{i};\bphi)
		\right]\bkappa.
  \label{F.6}
\end{align}
As long as we choose $\bkappa^{\ast}$ such that 
\begin{align}
	E \left\{
		\frac{\partial}{\partial \bphi}\theta_{\rm APSW, \gamma}(\bphi\mid  
		\blambda^{\ast},  {\bbeta}^{\ast}, \sigma^{\ast 2} ;\bmu^{\ast}, \bzeta^{\ast}, \nu^{\ast}, \bkappa^{\ast})\bigg|_{\bphi = \bphi^{\ast}}
	\right\} = \bzero,
	\end{align}
the linearization form is attained. 

In a nutshell, we can derive estimates for $\widehat{\bmu}$, $\widehat{\bzeta}$, $\widehat{\nu}$, and $\widehat{\bkappa}$ by solving the estimation equations, which result from equating the formula in \eqref{F.2} to $\mathbf{0}$ and that in \eqref{F.6} to $\mathbf{0}$.

\section{Computational Complexity}

The main computation involved in our proposed method is solving the estimating equation.  We apply the Newton-Raphson algorithm to solve the estimating equation. For each step, the computational complexity is $O(N^{3})$, where $N$ is the length of root. Recall that $\blambda \in \mathbb{R}^{L+1}$ and $\bphi \in \mathbb{R}^{p}$.  By the flowchart 1 in Fig 2 in our revised manuscript, in each iteration, the computational complexity for the augmented PSW estimator is $O(n \max\{p + L\} + p^{3} + L^{3} )$, where the first term is the complexity to compute the estimating equation and the last two terms are the complexity to solve the estimating equation. Therefore, suppose we set $K$ as the maximum iterations for the Newton-Raphson method, the final computational complexity is $O( K(n \max\{p + L\} + p^{3} + L^{3}) )$. In a similar fashion, the computational complexity for the robust augmented PSW estimator is also $O( K(n \max\{p + L\} + p^{3} + L^{3}) )$. As a result, the computational complexity is linear in $n$.

\bibliographystyle{chicago}
\bibliography{ref_refine}

\end{document}